\documentclass[sn-mathphys,Numbered]{sn-jnl}
\usepackage{graphicx}%
\usepackage{multirow}%
\usepackage{amsmath,amssymb,amsfonts,mathtools,slashed}%
\usepackage{amsthm}%
\usepackage{mathrsfs}%
\usepackage[title]{appendix}%
\usepackage{color, xcolor}%
\usepackage{textcomp}%
\usepackage{manyfoot}%
\usepackage{booktabs}%
\usepackage{algorithm}%
\usepackage{algorithmicx}%
\usepackage{algpseudocode}%
\usepackage{listings}%
\usepackage[normalem]{ulem}
\usepackage[utf8]{inputenc}

\definecolor{blue}{rgb}{0,0,0.5}
\definecolor{darkgreen}{RGB}{0,175,10}
\definecolor{brown}{RGB}{150,50,0}
\definecolor{AK}{rgb}{0,0,1.0}
\definecolor{YMW}{rgb}{0.1,0.6,0.1}

\newcommand{\ba}{\begin{eqnarray}}
\newcommand{\ea}{\end{eqnarray}}
\newcommand{\be}{\begin{equation}}
\newcommand{\ee}{\end{equation}}

\begin{document}

\title[A  guide to the QCD light-cone sum rules\\ for $b$-quark decays
]{
\hspace*{\fill} {\small SI-HEP-2023-25, P3H-23-087; RBI-ThPhys-2023-37 }\\[1cm]
A  guide to the  QCD light-cone sum rules\\ for $b$-quark decays }

\author[1]{\fnm{Alexander} \sur{Khodjamirian}}\email{khodjamirian@physik.uni-siegen.de}

\author[2]{\fnm{Bla\v zenka} \sur{Meli\'c}}\email{melic@irb.hr}

\author[3]{\fnm{Yu-Ming} \sur{Wang}}\email{wangyuming@nankai.edu.cn}

\affil[1]{\orgdiv{Center for Particle Physics Siegen (CPPS), Theoretische Physik 1},
\orgname{Universit\"at Siegen},
\orgaddress{
\street{Walter-Flex-Stra\ss e 3},
\postcode{D-57068} \city{Siegen}, \country{Germany}}}

\affil[2]{\orgdiv{Rudjer Bo\v skovi\'c Institute}, \orgname{
Division of Theoretical Physics}, \\
\orgaddress{\street{Bijeni\v cka 54},
\postcode{HR-10000} \city{Zagreb},   \country{Croatia}}}

\affil[3]{\orgdiv{School of Physics}, \orgname{Nankai University},
\orgaddress{\street{Weijin Road 94}, \\ \city{300071 Tianjin},~\country{China}}}

%\preprint{mmm}

%%==================================%%

\abstract{
We overview the current status and future perspectives
of the QCD-based method of light-cone sum rules. The two main
versions of these sum rules, using light-meson and $B$-meson distribution amplitudes
are introduced and the most important applications  of  the method
are discussed.  We also outline open problems and future perspectives
of this method.
}

%\keywords{Quantum chromodynamics, Sum rules, Bottom mesons, Form factors}



\maketitle

\section{Introduction}

The QCD-based method of light-cone sum rules (LCSRs) has already a long history.
It was originally developed in Refs.~\cite{Balitsky:1986st,Balitsky:1989ry}
for the process of a hyperon radiative decay and, independently, in Ref.~\cite{Chernyak:1990ag},
where LCSRs were for the first time applied to $B$-meson exclusive decays.
An application to the $D$-meson semileptonic decays followed then in Ref.~\cite{Ball:1991bs}.
The usefulness of this method for the calculation of $B$-meson transition form factors
was recognized already in the early applications \cite{Belyaev:1993wp,
Belyaev:1994zk,Ali:1993vd,Ball:1997rj}.

The LCSRs, in their different versions,
offer a realistic possibility to calculate various form factors
of   semileptonic or radiative $B$-meson decays in the large recoil region of the final light hadron.
The method can be extended to other hadronic amplitudes involved in $B$  physics, such as strong couplings and nonleptonic decay amplitudes.
It is also applicable to the decays containing  two light hadrons in the final state
and  involving   broad resonances. Neither the large recoil region, nor  resonance final states  are currently accessible to the lattice QCD calculations, and therefore LCSRs in several cases are the only available QCD-based method.

To describe the method in a nutshell: each  LCSR for a hadronic  transition form factor
starts from a correlator which is a vacuum-to-hadron matrix element  of two quark currents,
chosen to match a form factor we are interested in.  The correlator contains a quark-level transition current,
and a current  interpolating the other hadron involved in  this transition.
For this correlator,  an operator product expansion (OPE) near the light-cone is applied,
provided the external momenta of quark currents are far off-shell.
The OPE yields an analytical and factorizable expression for the correlator,
written in a form of a convolution of the calculable
hard-scattering kernel with  universal light-cone distribution amplitudes (DAs) of the hadron
entering the vacuum-to-hadron matrix element.
These DAs  absorb nonperturbative effects, and their parameters represent  a universal, process-independent input.
The factorized form of the OPE has its roots in  the early studies of the QCD asymptotics of
hadron form factors \cite{Chernyak:1977as, Efremov:1979qk,Farrar:1979aw,Lepage:1980fj}. There is a
crucial difference, though:  in the LCSR approach factorization is applied at the correlator level.
To access the form factor,  the tools of conventional QCD (SVZ) sum rule method
\cite{Shifman:1978bx}  are then applied to
the same correlator, including the hadronic dispersion relation and
the quark-hadron duality approximation. Combining these three main elements: the light-cone OPE,
dispersion relation and duality, leads to the resulting LCSR,
 which is an approximate analytical formula for the form factor
depending on a set of universal inputs.

Without going into further details in this brief introduction, we only mention
that the method of LCSRs allows us to calculate the so called
soft contribution to a form factor. It describes a transition
between the initial and final hadrons which is realized
as an overlap of two configurations with low-momenta spectator
quarks, hence, without a hard-gluon exchange. The soft contributions in a LCSR, being not accessible
in purely perturbative QCD-based methods, correspond to the
leading-order (LO), $O(\alpha_s^0)$  kernel of the OPE. The
hard-scattering part of the same form factor is then provided by the next-to-leading
(NLO) perturbative gluon corrections to the same kernel. As a result, the LCSR allows one to access the complete form factor within one method and input, albeit in an indirect way, via dispersion relation.

We emphasize
that the main goal  of this review is not to dwell on the derivational details of the LCSR method. These details  can be found in several reviews
(see e.g., Refs.~\cite{Braun:1997kw, Khodjamirian:1997lay, Colangelo:2000dp, Khodjamirian:2020btr}). We also do not cover here the
subtle aspects of twist expansion for nonlocal operators near light-cone
and the emergence of the corresponding light-cone distribution amplitudes.
These issues essential for LCSRs can be found in dedicated papers and reviews
and  the relevant ones  will be cited below.
Here we rather aim at presenting a short ``user guide" to  the LCSR applications
to various exclusive decays of $B$-meson, mostly to their form factors, but also to other
related hadronic quantities. We have made a selection of
all -- from our point of view -- important and useful papers  in this field, with a brief assessment of their
main results. We also tried to outline unsolved problems and future  perspectives of the  LCSR applications.

The rest of the review is organized as follows.  In Sections~\ref{sect:LCSRlight}
and \ref{sect:BDA} we overview the two main versions of the LCSR method
for $B$ meson transition form factors,  employing,
respectively, the light-meson and $B$-meson DAs.
In Section~\ref{sect:Bgamlnu}, we collect all important results concerning
the $B$ decays into photons and leptons. Section~\ref{sect:other} contains
a brief guide to other perspective applications of LCSRs for heavy hadron decays.
The concluding discussion is presented in Section~\ref{sect:outl}.

%%%%%%%%%%%%%%%%%%%%%%%%%%%%%%%%%%%%%%%%
\section{Sum rules  with light-meson distribution amplitudes}
%%%%%%%%%%%%%%%%%%%%%%%%%%%%%%%%%%%%%%%%
\label{sect:LCSRlight}
\subsection{Standard application: the $B\to \pi$ form factors  and  $B^*B\pi$ coupling }
\label{ssect:BpiLCSRlight}
The $B\to \pi$ form factors, generated in SM by the weak $b\to u $ transition,
were among  the first applications of LCSRs.
The vector and scalar form factors\footnote{We use standard definitions of these form factors, see, e.g.,
Ref.~\cite{Duplancic:2008ix}.} $f^+_{B\pi}$ and $f^0_{B\pi}$
determine the
weak semileptonic decay $\bar{B}^0\to \pi^+\ell \nu_\ell$.
To describe the rare $B\to \pi\ell^+\ell^-$ decay, generated by the
flavour-changing neutral current (FCNC) $b\to d$ transition, one needs, in addition to  $f^{+,0}_{B\pi}$,
also a  form factor $f_{B\pi}^T(q^2)$ of the tensor $b\to d$ quark current.

In the rest of this subsection, we briefly recall the
derivation  of LCSR for the most important form factor $f^+_{B\pi}$.
The underlying correlator is defined as a vacuum-to-pion matrix element of the time-ordered product of two currents:
\ba
F_{\mu}(p,q)&=&i\int d^4x ~e^{i q\cdot x}
\langle \pi^+(p)|T\left\{\bar{u}(x)\gamma_\mu b(x),
m_b\bar{b}(0)i\gamma_5 d(0)
\right\}|0\rangle
\nonumber\\
&=&F((p+q)^2,q^2)\,p_\mu +\widetilde{F}((p+q)^2, q^2)\,q_\mu\,,
\label{eq:corr}
\ea
where $q$ and $(p+q)$ are, respectively, the momenta of
the $b\to u$ weak current and of
the $B$-meson interpolating current.  Only the first invariant amplitude $F$ in the Lorentz-decomposition is relevant for $f^+_{B\pi}$. The  amplitude $\widetilde{F}$ is used to obtain a second LCSR, so that a linear combination of the two sum rules yields the scalar form factor. The tensor
form factor $f_{B\pi}^T$ is accessed by replacing the vector weak current in
the correlator by the tensor $b\to d$ transition current.  Ability to access different form factors by varying  currents or invariant amplitudes in the correlator reveals flexibility and universality of the LCSR method.

The correlator in Eq.~(\ref{eq:corr}) is calculated at the external momenta squared far below the open $b$-flavour thresholds, that is, at
\be
(p+q)^2\ll m_b^2, ~~q^2\ll m_b^2.
\label{eq:regions}
\ee
These conditions induce a strongly oscillating exponent in the correlator, retaining
 only  the  region near the light-cone $x^2\simeq 0$ in the
 four-coordinate integral. Hence,
the  product of currents can be expanded near $x^2=0$.
In particular, a highly virtual b-quark is
replaced with a light-cone expansion  \cite{Balitsky:1987bk}   of its propagator.
Contraction of $b$-quark fields generates
a perturbatively calculable kernel which is factorized from
the remaining long-distance part.
The resulting expression for the invariant amplitude $F$ (or similarly for $\tilde{F})$ has the following  schematic form:
%
%\begin{align}
\begin{eqnarray}
F((p+q)^2,q^2)&=&i \int d^4x\,e^{iqx}\Bigg\{
\big[ S_{0}(x^2,m_b,\mu)
+\alpha_s S_{1}(x^2,m_b,\mu) +\dots \big]
\nonumber\\
&\otimes& \langle \pi(p)\mid \bar{u}(x)
\Gamma d(0)\!\mid \!0 \rangle_{|_{\mu}}
%\hspace*{-1cm}
+\int_0^1 dv~\big[\tilde{S}_0(x^2,m_b,\mu,v) +\dots \big]
\nonumber\\
&\otimes& \langle\pi(p)\mid \bar{u}(x)G(vx)
\tilde{\Gamma}d(0)\}\mid 0\rangle_{|_{\mu}} +\dots \Bigg\}\, ,
\label{eq:OPEschem}
%\end{align}
\end{eqnarray}
where $S_0$ and $S_1$ are the LO and NLO parts of the perturbative kernel. They correspond,
respectively,
to the free $b$-quark propagator and $O(\alpha_s)$ perturbative gluon  corrections.
This kernel is convoluted with the  vacuum-to-pion matrix element
of the operator formed by a product of the $\bar{u}$ and $d$ quark fields at a near-light-cone separation,
with a generic Dirac structure $\Gamma$. The term with $\tilde{S}_0$  corresponds to
 a low-virtuality gluon emitted at a light-cone separation
from the $b$-quark propagator and  absorbed,
together with a quark-antiquark pair,
in the final pion state. The ellipsis in  Eq.~(\ref{eq:OPEschem})
denotes  higher-order terms, e.g., the NNLO, $O(\alpha^2_s)$ corrections to $S_0$,
NLO corrections to $\tilde S_0$ or the terms with two quarks  and two
antiquarks entering the vacuum-to-pion matrix element.
The diagrams and detailed expressions for this correlator
at NLO  are given in Ref.~\cite{Duplancic:2008ix},\
(see also the introductory review \cite{Khodjamirian:1997lay} for a description of derivation
at the LO level).

The scale  $\mu$ indicated in Eq.~(\ref{eq:OPEschem}) corresponds to a separation between  the
short-distance and long-distance parts in this convolution.
The optimal choice is $\mu\sim \sqrt{\Lambda m_b}$, where
an intermediate scale  $\Lambda\sim$ 1 GeV is parametrically larger
than $\Lambda_{\rm QCD}$, but at the same time does not scale with the heavy mass
$m_b$. This choice  guarantees that the average interval
$|x^2|\sim 1/\mu^2$, between the emission points of the
light-quark and gluon fields in the vacuum-to-pion matrix elements,
can still be considered small.
The subsequent transition of light constituents into an on-shell pion state includes all nonperturbative effects at energy-momenta below $\mu$.

The formula (\ref{eq:OPEschem}) reveals a general structure
of the light-cone OPE, emerging after expanding the correlator in terms of (i) quark-gluon coupling and (ii) multiplicity of light-quark and gluon fields entering the vacuum-to-pion matrix elements. The possibility to retain the lowest  terms in the expansions  (i) and (ii) is based on the suppression, respectively, due to extra powers of $\alpha_s$ (for which the scale $\mu$ is a natural choice) and due to inverse powers of $\mu$.

The complete  OPE of the correlator outlined in Eq.~(\ref{eq:OPEschem}) is in fact more involved. Within each vacuum-to-pion matrix element,  an additional expansion  emerges after expressing
these matrix elements  in terms of
the pion light-cone DAs with growing twists, starting from the
lowest twist-2  DA:
\begin{align}
\langle\pi(p)|\bar{u}(x)[x,0]\gamma_\mu\gamma_5 d(0)|0\rangle_{x^2\sim 0}=
-ip_\mu f_\pi\int_0^1du\,e^{iup \cdot x} \varphi_\pi (u) +O(x^2) + \dots
\label{eq:pi}
\end{align}
Here $u$ and $1-u$ are the fractions of the pion momentum $p$  distributed
among the two partons  (quark and antiquark) in the pion,
and $[x,0]$ is the gauge factor.
Two  similar vacuum-to-pion matrix elements emerging
in the light-cone expansion of the correlator contain (instead of $\gamma_{\mu}\gamma_5$) the Dirac-matrices $ \Gamma=i\gamma_5$
and $\Gamma=\sigma_{\mu\nu}$,  yielding, respectively, two additional DAs of the next-to-lowest twist-3. In addition,
the $O(x^2)$ terms in Eq.(\ref{eq:pi}) generate two more two-particle (quark-antiquark) DAs  of twist-4. The second hadronic matrix element in
Eq.~(\ref{eq:OPEschem})  is expressed via
three-particle (quark-antiquark-gluon) pion DAs, starting from twist 3. Currently, altogether five two-particle DAs with $t=2,3,4$ and five
three-particle DAs with $t=3,4$ are taken into account in the OPE.
Their  expressions, normalization parameters and other important features, such as the relations between  two and three-particle DAs due to QCD equations of motion, can be found in the most updated form in Ref.~\cite{Ball:2006wn}. A short
summary of these DAs and convenient defining formulas for the two- and three-particle vacuum-to-pion matrix elements are also given in Appendix A of Ref. \cite{Duplancic:2008ix}.

After substitution of DAs and the  coordinate integration,  the OPE  (\ref{eq:OPEschem})
for the invariant amplitude $F$
transforms into a sum  of separate DA contributions. In a compact generic form we have:
\ba
F_{OPE}((p+q)^2,q^2)=\sum\limits_{m=1}^{\infty}
\int {\cal D}u_{\{m\}}~
\sum\limits_{t\geq 2} \Bigg[T_0^{(m,t)}\big((p+q)^2,q^2,u_1,..u_m,\bar{m}_b,\mu\big)
\nonumber\\
+ \, \alpha_s T_1^{(m,t)}\big((p+q)^2,q^2,u_1,..u_m,\bar{m}_b,\mu\big)
+ O(\alpha_s^2)
\Bigg]
\phi_\pi^{(m,t)}(u_1,u_2,..u_m,\mu)\,,
\label{eq:OPE}
\ea
where $\phi_\pi^{(m,t)}$ is a pion DA with the light-parton multiplicity $m$ and twist $t$. This DA depends on the
pion momentum fractions $u_1,u_2,...u_m$, and we use the notation ${\cal D}u_{\{m\}}=  \displaystyle \int   \left ( \prod \limits_{i=1}^{m} du_i  \right ) \, \delta \left (1-
\sum\limits_{i=1}^m u_i \right )$  for the integration element.
In the above expression, the twist-2 DA defined in Eq.~(\ref{eq:pi}) corresponds
to $\phi^{(2,2)}(u,1-u)=f_\pi\varphi_{\pi}(u)$.
A nontrivial feature
of the light-cone OPE is the factorization proven at NLO for the leading
power twist-2 terms, so that the $\mu$ dependence
of the perturbative kernel $T_0^{(2,2)}+ \alpha_sT_1^{(2,2)} $ is compensated by the perturbative evolution of $\phi_\pi^{(2,2)}$,  well known as the ERBL evolution
\cite{Efremov:1979qk,Lepage:1980fj}.
Also the choice of the $\overline{\rm MS}$ scheme for the $b$-quark mass  $\overline{m}_b$ is
 justified for a correlator with a highly-virtual $b$-quark.
The currently known terms in Eq.(\ref{eq:OPE})  are  listed
in Table \ref{tab:OPEterms} with references to the papers
in which they have been computed and where  further important details can be found.
\begin{table}[h]
%\centering
%\scriptsize
%\small
\renewcommand{\arraystretch}{1.5}
%\resizebox{\columnwidth}{!}{
\begin{tabular}{c c c c}
\hline
multiplicity of DA & order in $\alpha_s$ & twist  & Ref.
%\multicolumn{4}{|c|}{... }
\\[-1mm]
\hline
\multirow{4}{*}{$m=2$} & LO & $t=2,3,4$  &\cite{Belyaev:1993wp}, \cite{Belyaev:1994zk}
\\[-1mm]
%\hline
& \multirow{2}{*}{NLO} & $t=2$ &\cite{Khodjamirian:1997ub}, \cite{Bagan:1997bp}\\[-1mm]
~& ~& $t=3$ & \cite{Ball:2001fp},  \cite{Duplancic:2008ix}
\\[-1mm]
& partial NNLO & $t=2$ & \cite{Bharucha:2012wy}
\\
\hline
$m=3$  & LO    & $t=3,4$ & \cite{Belyaev:1993wp}, \cite{Belyaev:1994zk}\\
\hline
$m=4$ & LO & $t=5,6$ & \cite{Rusov:2017chr}
\\[1mm]
\hline
\end{tabular}
\caption{ The calculated terms in the OPE expansion (\ref{eq:OPE}). The multiplicities
$m=2,3$ and $m=4$ correspond, respectively to the quark-antiquark, quark-antiquark-gluon DAs and to the diquark-antidiquark  DAs in the factorized approximation (quark condensate
$\otimes$ two-particle DA).}
\label{tab:OPEterms}
\end{table}

Finally, one more   expansion  implicitly present in the OPE
(\ref{eq:OPE}) concerns the dependence of each DA
on the fractions $u_i$ of the pion momenta shared between light degrees of freedom.
Here the formalism of conformal partial-wave expansion in QCD is used. Comprehensive reviews on that subject
can be found , e.g. in  Refs.~\cite{Braun:2003rp}, \cite{Braun:2022gzl}.
The conformal expansion represents a given pion DA in a form of series
in orthogonal polynomials. For the twist-2 DA it is the well familiar
series in Gegenbauer polynomials with multiplicatively renormalizable
coefficients:
\be
\varphi_\pi(u,\mu)=6u(1-u)\Big(1 +a_2^\pi(\mu)C_2^{3/2}(2u-1)
+a_4^\pi(\mu)C_4^{3/2}(2u-1) + \dots\Big)\,.
\label{eq:tw2DA}
\ee
The polynomial expansion  for this and other DAs, is usually taken up to the second conformal partial wave,  that is, retaining only $a^\pi_{2}$ and $a^\pi_4$ in the above formula.  A usual motivation is
that  the anomalous dimensions of polynomial coefficients grow with $n$,
suppressing terms with larger $n$, so that at
sufficiently large scale $\mu$, the DA  (\ref{eq:tw2DA}) is not
far from  its  asymptotic form, which is  $6u(1-u)$.
The  Gegenbauer moments  in $\varphi_\pi$ and similar  polynomial coefficients
for the higher-twist pion DAs, all taken at a reference scale $\mu=1.0$ GeV, represent
universal, process independent  inputs for the OPE.
Various methods to determine these parameters include
lattice QCD computation and two-point QCD sum rules. Another useful strategy applied to
determine the shape of the lowest twist-2 pion DA is to
fit to their measured values the pion transition form factor $\gamma\gamma^*\to \pi^0$  and  the pion electromagnetic form factor, both calculated from LCSRs
(see, respectively, e.g., Refs.~\cite{Agaev:2010aq,Agaev:2012tm, Mikhailov:2021znq}   and
Refs.~\cite{Khodjamirian:2011ub,Cheng:2020vwr}).

Having at hand the OPE result (\ref{eq:OPE}) for the invariant amplitude as an analytical function of the variable
$(p+q)^2$ at fixed $q^2$, one then applies the
standard tools of the conventional QCD sum rule method:

$\bullet$ equating this result to the hadronic dispersion relation in the variable $(p+q)^2$:
\be
F^{OPE}((p+q)^2,q^2)=\frac{2f_Bm_B^2f^+_{B\pi}(q^2)}{m_B^2-(p+q)^2}+...,
\label{eq:dispF}
\ee
where  the ellipsis indicates all heavier states with the $B$-meson quantum numbers, starting
from the lowest threshold at $m_{B^*}+m_\pi$.
Note that the hadronic dispersion relation is valid at any $(p+q)^2$,
whereas the relation (\ref{eq:dispF}) is only used in the region
(\ref{eq:regions}) of the OPE validity;

$\bullet$
using  quark-hadron semi-local duality approximation:
\ba
\int\limits_{(m_{B^*}+m_\pi)^2}^{\infty}\!\!\! ds\,\frac{\mbox{Im}F(s,q^2)}{s-(p+q)^2}=
\int\limits_{s_0^B}^{\infty}  ds\,\frac {[\mbox{Im}F(s,q^2)]_{OPE}}{s-(p+q)^2}\,,
\label{eq:dual}
\ea
where  the effective channel-specific threshold  $s_0^B$ is introduced;

 $\bullet$ subtracting the heavier state contributions from Eq.~(\ref{eq:dispF})
 with the  help of Eq.~(\ref{eq:dual}) and applying the Borel transform  $(p+q)^2\to M^2$.

The final form of LCSR is then obtained:
\be
m_B^2f_B  f_{B\pi}^+(q^2)e^{-m_B^2/M^2}=
\int\limits_{m_b^2}^{ s_0^B}  ds\, e^{-s/M^2}[\mbox{Im}F(s,q^2)]_{\rm OPE}\,.
\label{eq:lcsr}
\ee
This sum rule has three different sets of input parameters, ordered according to
their universality. The first one includes the $b$-quark mass and $\alpha_s$,
both determined with great accuracy. The second set contains  universal parameters of the pion DAs, including their normalizations and polynomial coefficients.
Finally, the third set includes parameters specific for the $B$-meson channel: (a)
the ranges of the scale $\mu$ and Borel parameter $M$; (b) the decay constant of $B$-meson
$f_B$,  that can be taken from lattice QCD average or, with a larger uncertainty,
from the two-point QCD sum rule calculation (see e.g., Ref.~\cite{Gelhausen:2013wia}); and (c) the threshold $s_0^B$ which is usually extracted from a
derivative sum rule obtained from LCSR. The latter procedure was systematically used in more recent analyses (see e.g., Ref.~\cite{SentitemsuImsong:2014plu}).

The latest numerical results for the $B\to \pi$
form factors at $q^2=0$ are given in the next subsection,
in Table~\ref{tab:resFF},
together with the form factors of all other $B$ transitions to  light pseudoscalar mesons. The standard $z$-expansion (here preferred is the BCL version
suggested in Ref.~\cite{Bourrely:2008za})
allows one to extrapolate the $B\to \pi$ form factors from the region of
LCSR validity,  typically at $q^2\leq 12-15 $ GeV$^2$  to larger $q^2$, that is, to
the low recoil region of the pion, where  this extrapolation can be compared with the lattice QCD predictions \footnote{ For brevity, we do not quote for comparison the lattice QCD results here, they can be found  in the updated and averaged form in Ref.~\cite{FlavourLatticeAveragingGroupFLAG:2021npn}, see also the review
on lattice QCD applications to $b$ quark physics   in this volume.}.

  The uncertainties quoted in
Table \ref{tab:resFF} are parametric and  they are
usually estimated in quadratures. A more advanced Bayesian analysis
of LCSRs  was initiated in Ref.~\cite{SentitemsuImsong:2014plu}.
There are however not many possibilities left to substantially decrease these
uncertainties in the future. Indeed, the still missing parts of OPE such as the
NLO corrections to the nonasymptotic twist-3 terms  and  to the twist-4 terms  are
expected to be very small. Also  the  factorizable twist-5, 6
contributions estimated in  Ref.~\cite{Rusov:2017chr} turned out to be negligible.
On the other hand, there is still a relatively large uncertainty
in the parameters of  the pion twist-2 DA.
Further efforts are desirable to improve the knowledge of this key element of LCSRs, combining
all available methods.

The most elusive uncertainty of systematic origin
in LCSRs remains the one related to the application of quark-hadron duality.
The attempts in the literature to quantify this uncertainty by varying the threshold parameter $s_0^B$ with the Borel scale or in any other way are only remedies, because these analyses still use duality as a basic assumption. In the future, one has to find alternative methods
to estimate the integrated hadronic spectral density of higher states
in the sum rules.

As an indirect way to assess the actual accuracy of LCSR predictions, we
suggest to increase  the accuracy of the $q^2$-shape in the measured
differential (binned) width of the $B\to \pi\ell\nu_\ell$ decays. This observable,
being directly proportional to the squared shape of the
form factor $f^+_{B\pi}(q^2)$, provides a direct test
of a QCD method, independent of the value of CKM parameter $V_{ub}$.

Finally, let us mention that straightforward byproducts of LCSRs with pion DAs considered in this subsection
are the analogous sum rules for the $D\to \pi$ form factors.
Switching to the charmed sector is straightforward, and demands
only a replacement of $b$ quark by the $c$ quark
in the correlator and a corresponding adjustment of all channel-specific inputs.
The last LCSR calculation  of these form factors in Ref.~\cite{Khodjamirian:2009ys}
deserves an update, e.g., including the twist 5,6 terms in the OPE. A comparison with experimental data on $D\to \pi\ell\nu_\ell$
semileptonic decay will allow one to tune the
universal input parameters of LCSRs and to further increase the accuracy of the $B\to \pi$  form factor determination.

The $B \to \pi$ form factor $f^+_{B\pi}(q^2)$ at large $q^2$ (near the zero  recoil of the pion), where the $B^{\ast}$-pole dominates, is determined by  the strong $B^*B\pi$ coupling. It is defined as the invariant constant parametrizing the hadronic matrix element
\be
\langle B^*(q) \pi(p)|B(p+q)\rangle =-g_{B^*\!B\pi}\,p^\mu\epsilon_\mu^{(B^*)}\,,
\label{eq:strong}
\ee
where $\epsilon_\mu^{(B^*)}$ is the polarization vector of $B^*$ meson.
The coupling $g_{B^*\!B\pi}$ is not measurable, since the $B^{\ast} \to B \pi$ decay is kinematically forbidden, contrary to its analog in charm sector, the  $D^{\ast}\to D\pi$ decay.

As originally suggested in Ref.~\cite{Belyaev:1994zk},
the coupling $g_{B^*B\pi}$ can be obtained from a LCSR, considering the same correlator (\ref{eq:corr}) and using its OPE. For the hadronic representation of the  invariant function $F(q^2, (p+q)^2)$ a double dispersion relation in both variables $p^2$ and $(p+q)^2$ should then be used:
\ba
F(q^2\!,(p\!+\!q)^2) = \frac{m_B^2 m_{B^*} f_B f_{B^*} \,g_{B^*B\pi} }{
(m_B^2\!-\!(p+q)^2)(m_{B^*}^2\!-\!q^2)} \!+\! \frac{1}{\pi^2}\!\!\int\!\!\!\!\int\!\! ds_2 ds_1 \frac{{\rm Im}_{s_1}{\rm Im}_{s_2} F(s_1, s_2) }{(s_2 \!-\!(p+q)^2)(s_1 \!-\! q^2)}
\,, \hspace{0.5 cm}
\label{eq:doubledisp}
\ea
where the lowest double-pole term contains the $B^*B\pi$
coupling multiplied by the decay constants  of pseudoscalar ($f_B$)
and vector ($f_{B^*}$) bottom mesons.
The duality approximation has to be defined for a
two-dimensional region in the $\{s_1,s_2\}$ plane,
adding an  uncertainty, related to the freedom to choose
the shape of that region, whereas the parametric accuracy of the OPE
is the same as in the LCSRs for $B\to \pi$ form factors.
Instead of a single Borel transform,
as  in Eq.~(\ref{eq:lcsr}), the double Borel
transform is then performed, removing all subtraction terms
that are not shown in Eq.~(\ref{eq:doubledisp}) for brevity, including single-variable dispersion integrals. Due to the approximate mass degeneracy of $B^*$ and $B$ mesons, usually the two Borel parameters, $ q^2 \to M_1^2$ and $(p+q)^2 \to M_2^2$, are taken equal, $M_1^2 = M_2^2 = 2 M^2$.

The strong coupling $g_{B^{\ast}B\pi}$ is then extracted from the sum rule:
\begin{align}
f_B \,f_{B^*}\,\,g_{B^*B\pi}
= \frac{1}{m_B^2 m_{B^*}}e^{\frac{m_{B}^2+m_{B^*}^2}{2M^2}}\!\!\!
\!\int\limits^{\Sigma(s_0)}\!\!\!\!\!\!\int\! ds_2 ds_1
e^{-\frac{s_2+s_1}{2M^2}}
\frac{1}{\pi^2}{\rm Im}_{s_1}{\rm Im}_{s_2} F_{OPE}(s_1, s_2)\,,
\label{eq:SR3}
\end{align}
%
%\begin{align}
%f_B f_{B^*}\,g_{B^*B\pi}
%= \frac{1}{m_B^2 m_{B^*}}\exp
%\left(\frac{m_{B}^2+m_{B^*}^2}{2M^2}\right)
%\bigg[\mathcal F^{\rm (LO)}(M^2,s_0) + {\alpha_s \, C_F \over 4 \pi} \,
%\mathcal F^{\rm (NLO)}(M^2,s_0)\bigg]\,,
%\label{eq:SR3}
%\end{align}
%where
%\begin{align}
%\mathcal F^{\rm (LO),(NLO)}(M^2,s_0)\equiv &
%\int\limits^{\infty}_{-\infty} \!d s_1
%\int\limits^{\infty}_{-\infty} \!d s_2\,
%\theta(2s_0-s_1-s_2)\,
%\exp\left(-\frac{s_1+s_2}{2\,M^2}\right)
%{\rm Im} F^{\rm (LO),(NLO)}(s_1,s_2) \,,
%\label{eq:FLO}
%\end{align}
%
where $\Sigma_0(s_0)$ indicates the duality region
parametrized with the effective threshold $s_0$.

In Ref.~\cite{Belyaev:1994zk} the LO result for this
sum rule at twist-4 accuracy was obtained.
The NLO correction to the twist-2 contribution was computed in Ref.~\cite{Khodjamirian:1999hb}.
The most recent and substantially
improved analysis of the LCSR (\ref{eq:SR3}) for the
$B^*B\pi$ coupling and (replacing $b\to c$ quark in the correlator)
for the $D^*D\pi$ coupling  is in Ref.~\cite{Khodjamirian:2020mlb},
where also the NLO twist-3 contributions were calculated.

In  Table \ref{tab:resSTRONG}, the results obtained in  Ref.~\cite{Khodjamirian:2020mlb} for the $B^{\ast}B\pi$ strong coupling
are displayed for two choices of the decay constants $f_B$ and $f_B^{\ast}$: from two-point sum rules, and from lattice QCD.
Also, two models for the leading twist-2 pion DA are considered.
Additional  details of this analysis and references relevant for the choice of the input parameters can be found in
Ref.~\cite{Khodjamirian:2020mlb}.

It is important to stress that the coupling $g_{B^*B\pi}$ is calculated from LCSR at a finite $b$-quark mass. Hence, replacing $b\to c$ in the sum rule provides also the charmed meson coupling $g_{D^*D\pi}$.
The infinitely heavy-quark limit of this coupling, known as the static coupling $\hat{g}$,
and serving as a key parameter
in the Heavy-Meson Chiral Perturbation Theory (HM$\chi$PT),
can also be  obtained from the same LCSR.
In Ref.~\cite{Khodjamirian:2020mlb} this limit was estimated,
together with the inverse heavy mass  correction combining the
couplings for both heavy mesons and fitting them to the parametrization:
\be
g_{H^* H \pi} = \frac{ 2 m_H \, \hat{g}}{f_{\pi}} \left (  1 + \frac{\delta}{m_H} \right )\,, ~~(H=D,B)\,.
\ee
%
%\be
%\lim_{m_Q\to \infty} g_{H^*\!H\pi}/(2m_H)= \hat{g}/f_\pi\,,
%\label{eq:heavy}
%\ee
%where the static coupling $\hat{g}$ does not depend on the heavy mass scale %and enters the HM$\chi$PT Lagrangian.
The results are given in Table \ref{tab:resSTRONG}.
%%%
\begin{table}[h]
\centering
%\scriptsize
\small
\setlength\tabcolsep{5pt}
\def\arraystretch{1.4}
\begin{tabular}{c c c c c c}
\toprule
$\varphi_\pi$ & decay constants & $g_{B^*B\pi}$ & $\hat{g}$ &$\delta$\,[GeV]\\
\midrule
\multirow{2}{*}{Model 1} & 2-point sum rule
& $24.1^{+4.5}_{-3.8}$
& $0.18^{+0.02}_{-0.03}$
& $3.28^{+0.62}_{-0.17}$
\\
& lattice QCD & $30.0^{+2.6}_{-2.4}$
& $0.30^{+0.02}_{-0.02}$
& $1.17^{+0.04}_{-0.04}$\\
\hline
\multirow{2}{*}{Model 2} & 2-point  sum rule
& $23.0^{+4.5}_{-3.8}$
& $0.17^{+0.03}_{-0.03}$
& $3.31^{+0.30}_{-0.01}$\\
& lattice QCD
& $28.6^{+3.0}_{-2.8}$
& $0.29^{+0.03}_{-0.03}$
& $1.18^{+0.00}_{-0.02}$\\
       %\hline
 &&&&&\\[-2mm]
       \botrule
\end{tabular}
\caption{LCSR  results \cite{Khodjamirian:2020mlb}  for the strong coupling of the bottom mesons for the two
choices of the decay constants and of the pion twist-2 DA.}
\label{tab:resSTRONG}
\end{table}
%%%

%%%%%%
\subsection{ The $B_{(s)}$-meson transitions to various light mesons}
\label{ssect:BKeta}
Analogously to the $B  \to \pi$ form factors, the other $B_{(s)}\to P $   form factors ($P=K,\eta,\eta'$)
can be obtained from LCSRs. These form factors are of a particular interest, for an alternative $V_{ub}$ determination, for various rare $B_{(s)} \to P \ell^+ \ell^-$ decays, and also for testing the factorization approximation in nonleptonic $B_{(s)}$ decays.

The initial correlator for a $B_{(s)} \to K$ transition is obtained from Eq.(\ref{eq:corr}), replacing the pion state with the kaon state and
making necessary changes in the interpolation and transition quark currents.
It is also important that  LCSRs allow for a complete account of $SU(3)_{fl}$-breaking effects,  originating from a nonvanishing $s$-quark mass and revealing themselves in differences between the kaon and pion DAs, starting from
the ratio $f_K/f_{\pi}$ as well as in the ratios of other hadronic parameters entering the sum rules, e.g., $m_{B_s}/m_B$ and $f_{B_s}/f_B$.

The lowest twist-2 DA of a kaon has an  expansion in Gegenbauer polynomials similar
to Eq.~(\ref{eq:tw2DA}), but including also the odd moments $a_{1,3,...}^K$,
in contrast to the pion DA, where due to the $G$-parity conservation, the odd moments vanish. The set of higher twist kaon DAs are worked out in Ref. \cite{Ball:2006wn}.

The $B_{(s)} \to K$ form factors were calculated from LCSRs first in Ref. \cite{Duplancic:2008tk} and more recently in Ref.~\cite{Khodjamirian:2017fxg}, with the same accuracy as for $B\to\pi$ form factors discussed in the previous subsection.
%%%%%%%% eta
The calculation of the $B_{(s)} \to \eta, \eta'$ transition form factors is somewhat more complicated due to the $\eta-\eta'$ mixing and a related  $U(1)_A$ QCD anomaly contribution to that mixing.
The first calculation of the $f^+_{B\eta}$ transition form factors at NLO level
for the leading twist-2 was done in Ref.~\cite{Ball:2004ye}.
 This calculation was further improved in Ref.~\cite{Ball:2007hb}, where also the $B \to \eta'$ transition was considered.
The $U(1)_A$ anomaly induces, in addition to flavour-singlet quark-antiquark DA the two-gluon DA  which contributes to the $B \to \eta,\eta'$ transitions, at NLO level, and has to be taken into account. This introduces additional uncertainty in the calculation since the coefficients in the Gegenbauer expansion of the twist-2 gluon DA are not known.
The most recent application of  LCSRs to all
the  $B_{(s)} \to \eta, \eta'$ form factors
(as well as to the $D_{(s)} \to \eta, \eta'$ form factors), at NLO and including two-gluon DAs is in Ref.~\cite{Duplancic:2015zna} (see also
Ref.~\cite{Offen:2013nma}).
The results for all form factors of the $B$-meson transitions to pseudoscalar mesons at $q^2=0$ are summarized in Table \ref{tab:resFF}.
%%%%EPJ table
\begin{table}[h]
\begin{tabular}{@{}llll@{}}
%\begin{tabular}{|c|c|c|c|}
%\scriptsize
\toprule
FF & $f^+(0) = f^0(0)$ & $f^T(0)$ & Ref.\\
\midrule
$B \to \pi$ & $ 0.297 \pm 0.030$ &$ 0.293 \pm 0.028$ &\cite{Leljak:2021vte}\\
 & $0.301 \pm 0.023$ &$0.273 \pm 0.021$ &\cite{Khodjamirian:2017fxg}\\
\hline
$B\to K$    &$0.395 \pm 0.033$& $0.381 \pm 0.027$ &\cite{Khodjamirian:2017fxg}\\
\hline
$B_s\to K$ & $0.364 \pm 0.026 $    &$0.394 \pm 0.023$&\cite{Bolognani:2023mcf}\\
& $0.336 \pm 0.023$ & $0.320 \pm 0.019$ &\cite{Khodjamirian:2017fxg}\\
\hline
&&&\\[-2.5mm]
$B\to \eta$&$0.168^{+0.041}_{-0.047}$ & $0.173^{+0.041}_{-0.035}$ &
%\cite{Duplancic:2015zna}
\\
%\hline
&&&\\[-2mm]
$B\to \eta'$&$0.130^{+0.036}_{-0.032}$&
$0.141^{+0.032}_{-0.030}$    &  \\
&&&\cite{Duplancic:2015zna}\\
$B_s\to \eta$  & $0.212^{+0.015}_{-0.013}$ &
$0.225^{+ 0.019}_{-0.014}$
& \\
&&&\\[-2mm]
$B_s\to \eta'$& $0.252^{+0.023}_{-0.020}$ &
$0.280^{+0.022}_{-0.016}$
& \\
\botrule
\end{tabular}
\caption{The most recent LCSR results for $B_{(s)} \to P$ form factors at $q^2=0$. Their full $q^2$ dependence is given in the corresponding papers.}
\label{tab:resFF}%
\end{table}

Extension of the LCSR method to the $B \to V$ form factors, where
$V = \rho, \omega, \phi, K^*$,
demands  a vacuum-to-$V$ correlator which otherwise has the same structure as the vacuum-to-$\pi$ correlator in Eq.~(\ref{eq:corr}).  Correspondingly,  the OPE is obtained in terms of vector meson distribution amplitudes. Importantly, these DAs are defined neglecting the widths of vector mesons.
As an example,  one  of the two twist-2 DAs of $\rho$-meson is defined as:
\be
\langle \rho(p)| \bar{u}(x)\sigma_{\mu\nu}[x,0] d(0)| 0\rangle=
-i f_\rho^\perp \big(\epsilon^{*(\rho)}_\mu p_\nu -p_\mu\epsilon^{*(\rho)}_\nu \big)
\int\limits_0^1 du\, e^{iu p\cdot x} \phi_\perp^{(\rho)}(u)\,,
\label{eq:phi_rho}
\ee
where $\epsilon^{(\rho)}$ is the polarization vector of $\rho$.
The DA $\phi_\perp^{(\rho)}$ corresponds to the
transversely polarized  $\rho$-meson and $f_\rho^\perp$ is the transverse decay constant. Note that the shape of this DA
is also determined by the Gegenbauer polynomial expansion. A comprehensive analysis of the twist-3 and twist-4 DAs
was done, respectively, in Refs.~\cite{Ball:1998sk} and \cite{Ball:1998ff}.

The very first LCSRs for $B \to V$ form factors were derived in Ref.~\cite{Ali:1993vd}, where the radiative $B \to V \gamma$ decays were considered.
The first application of this method to the semileptonic $B \to \rho \ell \nu$ decay was done, at LO level, in
Refs.~\cite{Ball:1996tb,Ball:1997rj}, where also the advantages of the LCSR approach over the  three-point QCD sum rules for the calculation of the heavy-to-light transition form factors were discussed in detail. A major update including
the NLO twist-2 contributions and extending the method to almost all  $B\to V$ channels was made in Ref.~\cite{Ball:1998kk}.
After that, in Ref.~\cite{Ball:2004rg} the analysis was improved by adding to OPE the NLO twist-3 terms and obtaining LCSRs also
for the $B \to \omega$ transition.
The most recent update of LCSRs with vector meson DAs for all $B_{(s)} \to V $ form factors, and with OPE based on the results of Ref.~\cite{Ball:2004rg},  can be found in Ref.~\cite{Bharucha:2015bzk}. In the latter paper, the form factors are also extrapolated to the whole semileptonic decay region after fitting them to the z-expansion.
Their results for the $B \to V$ transition form factors at $q^2=0$ are summarized in
Table \ref{tab:resFFV}.
%%

%%%%EPJ table B\to V
\begin{table}[h]
\begin{tabular}{@{}llll@{}}
%\begin{tabular}{|c|c|c/c/}
%\scriptsize
\toprule
FF          & $V(0)$            & $A_0(0)$          &  $T_1(0)=T_2(0)$  \\
   &                            & $A_1(0)$          &  $T_{23}(0)$  \\
   &                            & $A_{12}(0)$       &               \\
\midrule
$B\to \rho$ & 0.327 $\pm$ 0.031 & 0.356$\pm$ 0.042  & 0.272 $\pm$ 0.026\\
             &                   &0.262 $\pm$ 0.026  &  0.747 $\pm$ 0.076 \\
             &                   & 0.297 $\pm$ 0.035 &                        \\
\hline
$B\to \omega$& 0.304 $\pm$ 0.038 & 0.328 $\pm$ 0.048  &0.272 $\pm$ 0.026\\
             &                  &0.243 $\pm$ 0.031& 0.683 $\pm$ 0.090\\
             &                   &0.270 $\pm$ 0.040&                 \\
\hline
$B\to K^*$  & 0.341 $\pm$ 0.036  &0.356 $\pm$ 0.046&0.282 $\pm$ 0.031\\
            &                   &0.269 $\pm$ 0.029 &0.668 $\pm$ 0.083\\
            &                   &0.256 $\pm$ 0.033&                 \\
\hline
$B_s\to \phi$ &0.387 $\pm$ 0.033& 0.389 $\pm$ 0.045&0.309 $\pm$ 0.027\\
            &                   & 0.296 $\pm$ 0.027&0.676 $\pm$ 0.071\\
            &                   &0.246 $\pm$ 0.029&                 \\
\hline
$B_s\to K^*$ &0.296 $\pm$ 0.030 &0.314 $\pm$ 0.048&0.239 $\pm$ 0.024\\
            &                   &0.230 $\pm$ 0.025&0.597 $\pm$ 0.076\\
            &                   &0.229 $\pm$ 0.035&                 \\
\botrule
\end{tabular}
\caption{The LCSR results for $B_{(s)} \to V$ transition form factors at $q^2=0$ from Ref.~\cite{Bharucha:2015bzk}.}
\label{tab:resFFV}%
\end{table}

\subsection{Sum rules  with dipion distribution amplitudes}
%%%%%%
\label{ssect:B2piDA}

The LCSRs  for $B\to \rho,K^*$ form factors with the vector meson DAs considered in the previous subsection
are derived  neglecting the total widths of vector mesons. This is certainly a poor
approximation for such a broad resonances as $\rho(770)$ and $K^*(892)$. A more
comprehensive approach is to consider
the $B\to 2\pi $ and $B\to K\pi$ transitions, where $\rho$ and $K^*$ resonances are only a part, albeit dominant, of  the
dimeson $2\pi$ and $K\pi$ states, respectively. The phenomenology of $B\to 2\pi$ form factors
was studied in Ref.~\cite{Faller:2013dwa}  were one can find all necessary definitions (see also Ref.~\cite{Kang:2015ule}).

Here, as an example,  we consider the $\bar{B}^0\to \pi^+\pi^0 \ell \nu_\ell$ semileptonic transition in which
the $\bar{B}^0\to \rho^+$ form factors  are the resonance parts of the $\bar{B}^0\to \pi^+\pi^0$
form factors.  To avoid lengthy formulas, we take only the vector part of the weak $b\to u$ current, which yields a single form factor defined as:
 \be
 i\langle \pi^+(k_1) \pi^0(k_2)|\bar{u}\gamma^\mu b|\bar{B}^0(p)\rangle
    = -F_\perp (q^2,k^2,\zeta)\, \frac{4}{\sqrt{k^2 \lambda_B}} \,
i\epsilon^{\mu\alpha\beta\gamma} \, q_\alpha \, k_{1\beta} \, k_{2\gamma}\,,
\label{eq:BpipiFF}
\ee
where $k^2=(k_1+k_2)^2$ is the squared invariant mass  of the dipion state and $\zeta$ is an additional angular variable.
The $B\to 2\pi$  form factor is then expanded in partial waves corresponding to  angular momenta $\ell=1,3,5...$ of the dipion state, yielding a series in  Legendre polynomials in the angular
variable. Note that the partial waves with $\ell=0,2,4,...$ are forbidden for the $\pi^+\pi^0$ state on symmetry grounds.
The $B\to \rho$ form factor contributes only to  the $\ell=1$ component $F_\perp^{(\ell=1)} (q^2,k^2)$ of the $B\to 2\pi$ form factor  via  dispersion relation in the variable $p^2$ at fixed $q^2$:
\be
\frac{\sqrt{3}F_{\perp}^{(\ell=1)}(q^2,k^2)}{\sqrt{k^2}\sqrt{\lambda_B}}=
\frac{g_{\rho\pi\pi}}{m_\rho^2-k^2-im_\rho\Gamma_\rho(k^2)}\frac{V^{B\to \rho}(q^2)}{m_B+m_\rho}
+ ...\, ,
\label{eq:VrelFperp}
\ee
where $g_{\rho\pi\pi}$  is the strong $\rho\pi\pi$ coupling. In the above relation, the  energy-dependent width is inserted in the Breit-Wigner formula and contributions of excited states, such as the $\rho(1450)$ resonance, are
indicated by  the ellipsis. Although dispersion relation by itself is model-independent
and follows from analyticity and unitarity principles, a certain model-dependence is unavoidable in the resonance term. In any case, this relation is  the only realistic possibility
to extract the $B\to\rho$  form factor from the $B\to 2\pi$ form factor, taking into account not only
the $\rho$ width effect, but also the nonresonant background. The latter emerges
from the continuum and excited state contributions hidden under the ellipsis
in  Eq.~(\ref{eq:VrelFperp}).

To access  the $B\to 2\pi$ form factors directly, the method of LCSRs, resembling the one presented in the previous subsections, was suggested in  Ref.~\cite{Hambrock:2015aor}
 (see also Ref.~\cite{Cheng:2017sfk}).
 A correlator similar to
 Eq.~(\ref{eq:corr}) was used in which, instead of
a single  meson state, there is an on-shell
state of two pions with a variable invariant mass squared $k^2\geq 4m_\pi^2$. Applying light-cone OPE yields
an expression with a structure resembling Eq.~(\ref{eq:OPEschem}). But in this case, the perturbative kernels with a virtual $b$-quark are convoluted with the vacuum-to-dipion matrix elements of quark-antiquark or quark-antiquark-gluon operators. These matrix elements  are parametrized in terms of
a set of new objects, the dipion DAs.
The latter were introduced and used much earlier
\cite{Grozin:1983tt, Muller:1994ses,Diehl:1998dk, Polyakov:1998ze, Diehl:2000uv},
to describe hard exclusive processes with dimeson  states, such as $\gamma^*\gamma \to 2\pi$.

The dipion DAs  are also
classified by their twist and by the multiplicity of quark and gluon fields. Currently, only the most important
two-particle (quark-antiquark) DAs
of twist-2 are available. One of them  is  defined as:
\be
\langle \pi^+(k_1)\pi^0(k_2)|\bar{u}(x)\gamma_\mu[x,0] d(0) |0\rangle=
-\sqrt{2}k_\mu\int\limits_0^1du e^{iu(k\cdot x)}\Phi^{I=1}_{\parallel}(u,\zeta,k^2)\,,
\label{eq:dipionDA}
\ee
and the second one denoted as
$\Phi^{I=1}_{\perp}$
has a $\sigma_{\mu\nu}$ Dirac structure between
quark fields. The index $I=1$ reflects the isospin of
the $\pi^+\pi^0$  state. Full definitions and many important properties
of these DAs can be found in Ref.~\cite{Polyakov:1998ze}.
Both DAs undergo a double expansion in Legendre polynomials -- i.e. in partial waves
of the dipion state --  and in Gegenbauer
polynomials. The latter expansion reflects the momentum distribution between quark and antiquark
and has the same form as Eq.~(\ref{eq:tw2DA}).  The  coefficients of this double expansion
replace Gegenbauer moments $a_{2n}$ in Eq.~(\ref{eq:tw2DA})
and are complex valued functions of $k^2$, with  the phase reflecting
strong rescattering of the  final-state pions.
The local limit $x \to 0$ of the matrix element
(\ref{eq:dipionDA}) in the isospin
symmetry limit, is proportional to
the pion electromagnetic form factor in the timelike
region which is well measured. However, for the
second twist-2 DA  this normalization coincides with the timelike
form factor of the tensor current which is not directly
measurable.  A calculation of
this hadronic parameter together with a few lowest
Gegenbauer functions for both dipion DAs was only performed at small $k^2$ in
Ref.~\cite{Polyakov:1998td}, employing the instanton vacuum model of QCD.

The OPE of the correlator for the $\bar{B}\to \pi^+\pi^0$ form factors was obtained in Refs.~\cite{Hambrock:2015aor,Cheng:2017sfk} with  a twist-2
accuracy and in the LO. This expansion is  valid at sufficiently small dipion invariant masses, $k^2\ll m_b^2$ and, simultaneously,  in the large recoil region $q^2\ll m_b^2$.
The rest of the LCSR derivation is essentially  the same
as for  the sum rules with a single light-meson DAs discussed in Section~\ref{ssect:BpiLCSRlight}.
In particular, the same hadronic dispersion relation and duality in the channel of the $B$-meson interpolating current
are used. In the resulting LCSRs,  the
partial wave components $F_\perp^{(\ell)}(q^2,k^2)$ with $\ell =1,3,..$ are separated from each other.
As shown in Ref.~\cite{Hambrock:2015aor} in detail, the  sum rule for
$F_\perp^{(\ell=1)}$, together with its analogs  for the $B\to 2\pi$ form factors
of the axial weak current,  determine the proportion  of the $B\to \rho$  channel in the general $B\to 2\pi$ transition. In addition,  the  ratios  of the form factors with $\ell>1$
with respect to the lowest one with $\ell=1$ were estimated.

The method of LCSRs with dipion DAs has a considerable
potential for further improvement. To increase
the precision, one needs detailed studies
of the twist expansion for dipion DAs, including
the three-particle (quark-antiquark-gluon) DAs. On
the other hand, a better knowledge of Gegenbaeur functions
for the leading twist-2 DA is necessary. A possibility to gain some information
on these universal functions from the  $D\to 2\pi \ell \nu$ decays (using the $b\to c$ replacement in the LCSRs)
is currently being studied \cite{RKAKGTX}. In the future,
in order to extend this method to other important form factors, DAs for the different states of two light pseudoscalar mesons should also be studied. Most important are  the dipion states with the spin-parities $J^P=0^+, 2^+$
relevant for $B\to \pi^+\pi^-$ form factors, as well as the $K\pi$ and $K\bar{K}$ states needed for  the $B\to K^*$ and $B\to \phi$ form factors, respectively.
In this respect, let us mention an earlier
paper \cite{Meissner:2013hya} where the form factors of $B$
transitions to the scalar  dimeson ($K\pi$ and $2\pi$)  were obtained from LCSRs.

%%%%%%    Vub
\subsection{Uses of LCSR form factors in the Standard Model tests}
%%%%%%
\label{ssect:BSM}
The $B$-meson transition form factors obtained from the LCSRs with light-meson DAs are mainly used to determine the modulus of the CKM matrix element $V_{ub}$
from  the data on exclusive semileptonic $b \to u \ell\nu_\ell$ decays, predominantly using the
 $B \to \pi\ell\nu_\ell$ decay, but also the $B_s \to K \ell\nu_\ell$, $B \to \rho \ell\nu_\ell$ and $B \to \omega\ell\nu_\ell$ decays.

One way to extract $|V_{ub}|$ is to use the  LCSR result for the $B\to\pi$ form factor $f^+_{B\pi}(q^2)$
and integrate  the  predicted differential $B\to \pi\ell\nu_\ell$ width  ($\ell=e,\mu$) over the LCSR validity region
$0<q^2<q^2_{max}$ (see e.g. Refs.~\cite{SentitemsuImsong:2014plu,Khodjamirian:2017fxg}).
The other way is  to extrapolate the form factor up to the
zero recoil point of the pion , $q^2=(m_B-m_\pi)^2$ using the $z$-parametrization, e.g. the one in
Ref.~\cite{Bourrely:2008za}. A combined fit to both LCSR and lattice QCD predictions can also be performed,
to achieve the theoretically most accurate form factor in the whole semileptonic region of $q^2$.

The recent determination of $|V_{ub}|$ using the combined approach to the  $B \to \pi$  form factors was performed in Ref.~\cite{Leljak:2021vte}.
An independent  extraction of $|V_{ub}|$ from $B\to \rho(\omega)\ell\nu_\ell$ decays
was performed in Ref.~\cite{Bernlochner:2021rel} using the LCSR form factors from Ref.~\cite{Bharucha:2015bzk}.
In Ref.~\cite{Leljak:2023gna}, the channels with vector mesons were added to the combined
analysis of the $B\to \pi\ell \nu_\ell$ decay, employing an advanced
statistical tool \cite{EOSAuthors:2021xpv}. Separate process-specific $|V_{ub}|$ values were obtained:
%
%\ba
$|V_{ub}|_{B \to \pi} = (3.79 \pm 0.15) \cdot 10^{-3}$,~
$|V_{ub}|_{B \to \rho} = (2.92^{+0.28}_{-0.25})\cdot 10^{-3}$,
$|V_{ub}|_{B \to \omega}  =  (3.00^{+0.38}_{-0.32}) \cdot 10^{-3} $,
%\qquad
%\ea
with an overall average:
\be
|V_{ub}| = (3.59 ^{+ 0.13}_ {- 0.12}) \cdot 10^{-3}\,.
\label{eq:Vubfin}
\ee

The observed difference between the $|V_{ub}|$ values extracted using $B\to \pi$ and
$B\to \rho(\omega) $  form factors demands further investigation, in particular, an
update of LCSRs for $B\to V$ form factors is desirable. The point is that
the twist-3 NLO contributions to LCSRs for $B\to V$ form factors are not available in analytical form and should be completely recalculated and reassessed \cite{rhoCOLL}.
It would also be useful to quantify
the nonresonant background for $B\to \rho (\omega)\ell\nu_\ell$  decays, along the lines presented in Ref.~\cite{Hambrock:2015aor} as already discussed in Section~\ref{ssect:B2piDA}.

Returning to the CKM matrix elements, quite
recently, the combined analysis of the $B_s \to K$ form factors was done
in  Ref.~\cite{Bolognani:2023mcf} and  used to extract the ratio
 $|V_{ub}/V_{cb}|$ from the data on the $B_s\to K \ell \nu_\ell $
 and $B_s\to D_s \ell \nu_\ell $ decays. A similar  analysis was  also done in  Ref.~\cite{Biswas:2022yvh}.

The $B\to K^{*}$ form factors obtained from LCSRs
were also extensively used in the exploration of various observables in the $B \to K^{*} \ell^+\ell^-$ and $B\to K^*\gamma$ FCNC decays.
Already in Ref.~\cite{Altmannshofer:2008dz} and after that in  Ref.~\cite{Bharucha:2015bzk} these observables were extended from SM  to various models of new physics.
An independent determination
of Wolfenstein parameters of CKM matrix  from
a combination of observables in $B_{(s)} \to P \ell^+ \ell^-$ decays was suggested
in Ref.~\cite{Khodjamirian:2017fxg}.

In the current tests of the lepton flavour universality (LFU) in semileptonic $B$ decays, the LCSR results for the form factors, combined with the lattice QCD results, were also employed. The main goal was to obtain predictions of LFU ratios, such as $R_{\pi}$ in Ref.~\cite{Leljak:2021vte} or $R_{\rho}$ and $R_{\omega}$ in Ref.~\cite{Bernlochner:2021rel}, where also a possible influence of new physics on the polarizations and asymmetries in $B \to (\rho, \omega) \ell \nu_\ell$ decays was  examined. Note that for LFU tests involving $B \to P \tau \nu_{\tau}$ decays, the scalar form factors $f_{BP}^0(q^2)$ are essential.  The latter ones, as well as the tensor form factors $f_{BP}^T(q^2)$, are not always available from the lattice QCD, enhancing even more the importance of their LCSR calculations.

Based on the  form factors extracted from LCSRs, in Ref.~\cite{Leljak:2023gna} also the effects of possible new physics were analysed in the framework of the Weak Effective Theory (WET). Similarly, the LCSR results for $B \to \pi, \rho, \omega$ form factors from Refs~\cite{Duplancic:2008ix} and \cite{Bharucha:2015bzk} were used in Ref.~\cite{Greljo:2023bab} to examine new physics interpretation in the Standard Model Effective Field Theory (SMEFT).

%%%%%%%%%%%%%%%%%%%%%%%%%%%%%%%%%%%%%%%%
\section{Sum rules with $B$-meson distribution amplitudes}
%%%%%%%%%%%%%%%%%%%%%%%%%%%%%%%%%%%%%%%%
\label{sect:BDA}

\subsection{An alternative method for
the $B$-meson form factors}
\label{ssect:LCSRBDAs}
The underlying idea of this version of LCSRs suggested in Ref.~\cite{Khodjamirian:2005ea} and developed further in
Ref.~\cite{Khodjamirian:2006st} was to swap the meson state and interpolating
current in the initial correlator. E.g., for the $B\to\pi$ transition one has  to consider, instead of Eq.~(\ref{eq:corr}), the $B$-to-vacuum
correlator with the pion interpolating current:
\begin{equation}
F_{\mu\nu}^{(B)}(p,q)= i\int d^4x ~e^{i p\cdot x}
\langle 0|T\left\{\bar{d}(x)\gamma_\mu\gamma_5 u(x),
\bar{u}(0)\gamma_\nu b(0)\right\}|\bar{B}(p+q)\rangle\,.
\label{eq:corrB}
\end{equation}
This opens up
a possibility to obtain form factors of $B$-meson transitions
into any light or charmed meson by just switching from one interpolating
current to another, and, correspondingly, applying quark-hadron duality in that  channel. There
is no need to introduce a separate set of DAs for a particular light meson.
New nonperturbative objects that emerge in the OPE of the correlator
(\ref{eq:corrB}),
after contracting the
virtual $u$-quark fields,  are the universal DAs of $B$ meson. However,
these DAs can only be systematically defined in the infinite heavy-quark mass limit,
in the framework of heavy-quark effective theory (HQET). Hence, we have to replace
the $B$-meson state in the correlator by a state $B_v$ with a definite velocity
$v=(p+q)/m_B$, and the $b$-quark field by an effective HQET field $h_v$.

The leading twist-2 and subleading twist-3 $B$-meson DAs
\footnote{Note that the concept  of twist for the $B$ meson DAs
introduced and explained in detail in Ref.~\cite{Braun:2017liq}
differs  from  the one for the light-meson DAs.},
 denoted, respectively as
$\phi_{+}^B$ and $\phi_{-}^B$ have been defined originally in
Ref.~\cite{Grozin:1996pq} (see also Ref.~\cite{Beneke:2000wa} and the review
\cite{Grozin:2005iz}):
\begin{eqnarray}
&& \langle 0|\bar{d}_{\beta}(x)[x,\!0] h_{v, \, \alpha}(0)
|\bar{B}_v\rangle
\nonumber \\
%&&
&& \hspace{0.2 cm} \!=\! -\frac{if_B m_B}{4}\!\!\!\int\limits_0^\infty \!\!
d\omega\, e^{-i\omega v\cdot x}
\left [ (1 \!+\!\slash \!\! \!v)\!
\left \{ \!\phi^B_+(\omega) \!-\!
\frac{\phi_+^B(\omega)\! -\!\phi_-^B(\omega)}{2 v\cdot x}\slash\!\!\! x \!\right \}\!\gamma_5 \right ]_{\alpha \beta},
\label{eq-BDAdef}
\end{eqnarray}
where the normalization constant is taken equal to the physical $B$-meson decay constant (as usually done in LCSRs at LO considered in this subsection).
The HQET distribution amplitudes also serve as an indispensable  ingredient for the theory descriptions
of exclusive $B$-meson decay matrix elements in the QCD factorization framework  \cite{Beneke:2000ry}.

The  LCSR derivation starting from the HQET limit of the correlator
(\ref{eq:corrB}) basically
repeats the procedure described in Section~\ref{sect:LCSRlight}. Hadronic dispersion
relation of the correlator in the pion channel (in the invariant variable $p^2$)
is now used, and one has to isolate the pion pole contribution with the
help of duality approximation, introducing the effective threshold $s_0^\pi$.
The resulting sum rule for the $B\to\pi$ vector form factor has a surprisingly simple expression in the lowest
twist-3 approximation, after Borel transformation $p^2\to M^2$:
\ba
f^+_{B\pi}(0)=
\frac{f_B }{ f_\pi\,m_B}
    \int\limits _0^{s_0^\pi} ds e^{-s/M^2}
    \phi^B_-(s/m_B)\,.
\label{eq-fplBpi0}
\ea
The r.h.s. of this sum rule becomes significantly more involved after adding higher-twist (power suppressed) contributions, including those from the quark-antiquark-gluon B-meson DAs.
The first calculations for $B\to \pi$ and $B\to K, \rho,K^*$ form factors in Ref.~\cite{Khodjamirian:2006st} were superseded by
more recent results for  these  form factors in
Refs.~\cite{Lu:2018cfc} and \cite{Gubernari:2018wyi}. In both analyses  the set of higher-twist DAs
worked out in  Ref.~\cite{Braun:2017liq} was taken into account. Note that there is an important difference in the achieved accuracy:
in   Ref.~\cite{Lu:2018cfc} both  two- and three-particle DAs were included
(the latter up to twist-six level) whereas in Ref.~\cite{Gubernari:2018wyi} only the two-particle
twist-five contributions were taken into account, which is
not consistent from the point of view of HQET equations of motion \cite{Braun:2017liq}
(see Refs.~\cite{Beneke:2018wjp,Gao:2019lta} for further discussions).

The  LCSRs described in this section were obtained
in the LO, that is, at the zeroth order in $\alpha_s$, albeit with the corrections beyond
the leading power (LP).
Therefore, renormalization and scale dependence for $B$-meson DAs,
with their specific evolution discovered  in Ref.~\cite{Lange:2003ff} and discussed in some detail
in the next subsection are not fully used here. As a result, the accuracy of OPE in these sum rules   is still
less than for the  LCSRs with light-meson DAs. Also
the key parameter of leading-twist DA -- the inverse moment
$1/\lambda_B=\int_0^\infty d\omega \phi_{+}^B(\omega)/\omega$ --
is not yet determined with a sufficient accuracy, in particular, this moment is not yet accessible in lattice QCD (see, however, \cite{Wang:2019msf} for interesting  discussions in this respect).
The input interval of this parameter employed in LCSRs for $B_{(s)}$ meson is usually  taken from the two-point QCD sum rules worked out in \cite{Braun:2003wx} (see also Ref.~\cite{Khodjamirian:2020hob} and  other independent
estimates in \cite{Wang:2015vgv,Feldmann:2023aml}). The behaviour  of the DAs at
$\omega\to 0 $  is model-independent \cite{Braun:2017liq}. The actual form
of these DAs at large $\omega$ is unimportant at tree level, since in the sum rules such as Eq.~(\ref{eq-fplBpi0}) the duality interval cuts
off the region $\omega>s_0^\pi/m_B$. One popular choice of $\omega$-dependence
is the exponential model from Ref.~\cite{Grozin:1996pq}
(see  also Refs. ~\cite{Lange:2003ff,Lee:2005gza,Feldmann:2014ika} ).

The two additional parameters determining all higher twist DAs in this model  are $\lambda_B$ and $\lambda_H$,
parametrizing   the quark-antiquark-gluon $B$-to-vacuum matrix elements in HQET.
Their estimates from independent two-point sum rules \cite{
Nishikawa:2011qk,Rahimi:2020zzo}
still yield   large uncertainty intervals.

Numerically, the form factors
of $B$ meson transitions to light mesons obtained from the LCSRs with
$B$ meson DAs (see the Tables in the next subsection) agree with the ones calculated from the sum rules  with light-meson DAs,
but only within their larger uncertainties.
Apart from narrowing down the uncertainty intervals of the inputs, there are two important questions to be addressed
for a further improvement of this version of LCSRs.
The first question, already raised in Ref.~\cite{Khodjamirian:2006st}, is the size of inverse heavy-quark mass  effects which are
implicitly neglected when one starts from the correlation function (\ref{eq:corrB}) in full QCD
with an on-shell $B$-meson state and then takes the HQET limit of that state.
A pragmatic argument in favour of smallness of such correction is a relatively good agreement between
form factors obtained with  both LCSR methods with light and
$B$-meson DAs. The second question  is the size of NLO, $O(\alpha_s)$ corrections to the correlator (\ref{eq:corrB}).
The answer is found using an alternative  formulation of this method to be discussed below.

Finally, we would like to comment on the possibility
to extend the method of LCSR with $B$-meson DAs
to $D$-meson semileptonic form factors, introducing the $D$-meson DAs.
This however implies using HQET for the $c$-quark and $D$-meson,
hence, considerably limiting  the accuracy  in phenomenological applications
to the $D$-meson  decays.

 \subsection{
 Accessing the next-to-leading order with  SCET  }
\label{ssect:Blight}

The method of LCSRs with $B$-meson DAs
was independently formulated \cite{DeFazio:2005dx,DeFazio:2007hw}
in the framework of
soft-collinear effective theory (SCET) (see \cite{Beneke:2002ph} where this theory
is introduced in the context of heavy-to-light transitions).
To derive the SCET sum rules for our standard example -- the $B\to \pi$ form factors -- the following
definition of the correlator is used (see e.g. \cite{Wang:2015vgv}):
\begin{eqnarray}
{\cal F}_\mu^{(B)}(n \cdot p, \bar n \cdot p)  =  \int d^4 x \, e^{i p \cdot x} \langle 0 | {\rm T} \left \{
		\bar d(x)\, \slashed {n} \gamma_5 \,u(x),\, \bar u(0)\, \gamma_\mu \,b(0) \right  \} | \bar B (p_B) \rangle
\nonumber\\
={\cal F}^{(B)}(n \cdot p, \bar n \cdot p)\, n_\mu + \tilde {\cal F}^{(B)} (n \cdot p, \bar n \cdot p)\, \bar n_\mu\,,
\label{eq:corrSCET}
\end{eqnarray}
where in the $B$-meson rest frame the two light-cone vectors $n_{\mu}$ and $\bar n_{\mu}$ are introduced, such that
$n \cdot v = \bar n \cdot v =1$, $v_\perp=0$ and $n \cdot \bar n =2$,
and a power-counting scheme for the four-momentum of the pion
interpolation current is employed:
\begin{eqnarray}
n \cdot p \sim {\cal O}(m_b) \,, \qquad \bar n \cdot p \sim {\cal O}(\Lambda_{\rm QCD})  \,.
\label{power-counting scheme}
\end{eqnarray}
In this scheme, a generic momentum $P_{\mu} \equiv (n \cdot P, \bar n \cdot P, P_{\perp} )$ is split into the three different momentum modes: hard (h), hard-collinear (hc) and soft (s), with the scaling behavior,
respectively,
%\begin{eqnarray}
$P_{h, \,  \mu} \sim {\cal O}(1, 1, 1)$,
$P_{hc, \, \mu} \sim {\cal O}(1, \lambda, \lambda^{1/2})$,
$P_{s, \, \mu} \sim {\cal O}(\lambda, \lambda, \lambda)$.
%\end{eqnarray}
The expansion parameter $\lambda$ scales as $\Lambda / m_b$ where $\Lambda$ is a typical hadronic scale.
The momentum transfer $q$ in the large and intermediate recoil region of the pion (accessible to LCSRs) belongs to the hard or hard-collinear mode.

The OPE for a  $B$-to-vacuum correlation function such as the one in Eq.~(\ref{eq:corrSCET}) is
directly calculated in terms of SCET diagrams.
In the leading, zeroth order in $\alpha_s$ the resulting LCSRs in SCET are fully equivalent to the sum rules discussed in the previous subsection. The advantages of SCET are revealed by going
beyond the LO approximation, starting from the NLO with one-loop gluon radiative corrections taken into account in the correlator (\ref{eq:corrSCET}).
Application of SCET  enables one to effectively resum large logarithms emerging in the heavy quark limit and to extend the factorization of the OPE expression to the NLO level.
In practice, achieving NLO accuracy is  accomplished by invoking a matching procedure,
with a two-step transition ${\rm QCD} \to {\rm SCET_{I}} \to {\rm SCET_{II}}$. The details of these ``embedded" effective theories
and their correspondence to the one-loop diagrams are
discussed already in Refs.~\cite{DeFazio:2005dx,DeFazio:2007hw}. Further development of both  calculational
and conceptual aspects of this procedure can be found in  Refs.~\cite{Wang:2015vgv,Wang:2017jow,Lu:2018cfc,Gao:2019lta,Gao:2021sav,Cui:2022zwm}.
Here we only quote the schematic form of the LP factorization formula for the particular invariant amplitude ${\cal F}^{(B)}$   \cite{Wang:2015vgv,Cui:2022zwm}
\begin{equation}
{\cal F}^{(B)}\! = \!{\tilde f}_B(\mu) \, m_B \sum \limits_{k=\pm} \,
{\cal C}^{(k)}(n \!\cdot\! p, \mu)\!\! \int_0^{\infty} \!\!{d \omega \over \omega- \bar n\! \cdot\! p - i 0}~
{\cal J}^{(k)}\left({\mu^2 \over n \!\cdot\! p \, \omega},{\omega \over \bar n \!\cdot\! p}\right) \
\phi^B_{k}(\omega,\mu)\,,
\label{eq:factB}
\end{equation}
where ${\tilde f}_B(\mu)$ is the HQET decay constant of $B$-meson. In this formula, ${\cal C}^{(k)}$ and  ${\cal J}^{(k)}$
are, respectively, the hard  and jet functions, both stemming from the matching procedure. The $B$-meson DAs absorbing soft contributions
in this formula are, up to the scale dependence, the same as the ones defined in the previous subsection.

Achieving the NLO level in LCSRs demands  including into the computational scheme
the scale dependence of the $B$-meson DAs, stemming from the evolution equations.
In this direction, there was a lot of progress
 in recent years. The renormalization-group (RG) evolution equation for  $\phi^B_{+}(\omega,\mu)$ determined at one loop \cite{Lange:2003ff},
was upgraded to the two loops \cite{Braun:2019wyx,Liu:2020ydl}. An explicit solution of this evolution equation was constructed with an integral transform method
in Ref.~\cite{Bell:2013tfa} and, independently, with the conformal symmetry technique in Ref.~\cite{Braun:2014owa}
(see also Ref.~\cite{Galda:2022dhp} for the analytical solution of the two-loop RG equation).
In addition,  the one-loop evolution equation for the twist-three distribution amplitude  $\phi^B_{-}(\omega,\mu)$
was constructed within the so-called Wandzura-Wilczek approximation, also
including the RG mixing effect generated by the non-vanishing light-quark mass  \cite{Bell:2008er}.
Then, in Refs.~\cite{Descotes-Genon:2009jif,Braun:2017liq},  the three-particle higher-twist  contribution $\Phi_3(\omega_1, \omega_2, \mu)$ was included.
Furthermore, an explicit solution to the RG equation for $\phi^B_{-}(\omega,\mu)$ was found
\cite{Braun:2015pha} by exploring the ``hidden" symmetries  of the evolution kernel of $\Phi_3$ in the large $N_c$ limit.

Employing the RG equations for the HQET $B$-meson DAs and for the perturbative matching functions, it became possible \cite{Wang:2015vgv,Lu:2018cfc,Cui:2022zwm}
to carry out an all-order resummation of the parametrically enhanced logarithms appearing in the soft-collinear factorization
formulae of the  $B$-meson-to-vacuum correlation functions.
 After the resummation  and all subsequent standard steps in the derivation are done,
the LCSR in SCET in the next-to-leading-logarithmic (NLL) approximation is obtained,
valid in the large recoil region.
For the vector  $B \to P $  form factor $f_{BP}^+$,
($P=\pi,K$) it has the following form:
\begin{eqnarray}
 f_{P} \,\, {\rm exp} \left [- {m_{P}^2 \over n \cdot p \,\, \omega_M} \right ] \,
\frac{n \cdot p} {m_B} \, f_{B P}^{+}(q^2)
=   \left [ \hat{U}_2(\mu_{h2}, \mu) \, \mathcal{F}_B(\mu_{h2}) \right ]
\,\, \int_0^{\omega_s} \,\, d \omega^{\prime} \, e^{-\omega^{\prime}/\omega_M} &&
\nonumber \\
\bigg \{ \widetilde{\bf \Phi}^{B, \, \rm {eff}}_{+} (\omega^{\prime}, \mu)
+  \, \left [  \hat{U}_1(n \cdot p, \mu_{h1}, \mu) \,\, \widetilde{\cal{C}}^{(-)}(n \cdot p, \mu_{h1}) \right ]\,
\widetilde{\bf \Phi}^{B, \, \rm {eff}}_{-} (\omega^{\prime}, \mu)  &&
\nonumber \\
+  \, { n \cdot p - m_B \over m_B} \,
\left [{\bf \Phi}^{B, \, \rm {eff}}_{+} (\omega^{\prime}, \mu)
+ {\cal C}^{(-)}(n \cdot p, \mu_{h1}) \, {\bf \Phi}^{B, \, \rm {eff}}_{-} (\omega^{\prime}, \mu)   \right ]  \bigg \}
+ {\cal O}(\alpha_s^2, \, \Lambda/m_b)\,,
&&
\label{eq:NLL}
\end{eqnarray}
where $\omega_M$ is the Borel parameter, $\hat{U}_{1, \, 2}$
are the evolution functions for
for the RG improved hard functions ${\cal C}^{(-)}$ and $\widetilde{\cal{C}}^{(-)}$.
The functions ${\bf \Phi}_{B, \, \rm {eff}}^{\pm}$ and $\widetilde{{\bf \Phi}}_{B, \, \rm {eff}}^{\pm}$  are the effective ``distribution amplitudes" which encode both the hard-collinear and soft strong interaction dynamics.
The explicit expressions for all these functions and further explanatory discussions
concerning the formula (\ref{eq:NLL}) can be found in Ref.~\cite{Cui:2022zwm}.
In the same  work, a  comprehensive analysis of the power-suppressed contributions to the $B \to \pi$ form factors in the combined LCSR/SCET framework
 was accomplished, applying non-trivial operator identities due to the HQET equations of motion \cite{Kawamura:2001jm,Braun:2017liq}. Four different sources of the
subleading power corrections at tree level were included:
i) the higher-order terms from  heavy-quark expansion of the hard-collinear quark propagator,
ii) the subleading power corrections from the effective matrix element of the ${\rm SCET_I}$ weak current, iii) the higher-twist corrections from the two-particle and three-particle HQET distribution amplitudes  at twist-six,
and (iv) the four-particle twist-five and twist-six contributions in the factorization approximation.

Extension of these analyses to NLO (${\cal O}(\alpha_s)$) accuracy is currently in progress \cite{BtoPNLP}, bearing in mind that the higher-twist three-particle HQET DAs also generate LP contributions to the heavy-to-light form factors at ${\cal O} (\alpha_s)$ \cite{Beneke:2002ph}.
The one-loop QCD corrections to the short-distance matching coefficients in the SCET sum rules for the $B_{d, s} \to \pi, K$ form factors at the LP accuracy can shift the corresponding tree-level predictions by an amount of ${\cal O} (30 \%)$ numerically \cite{Cui:2022zwm}. Moreover,
after these  sum rules are improved by the  NLL resummation,  the perturbative uncertainties from varying the hard and hard-collinear  matching scales are considerably pinned down.

The SCET sum rule method with the $B$-meson  DAs  is easily
extendable to  the semileptonic $B \to V$ form factors (with $V=\rho, \, \omega, K^{\ast}$) at large hadronic recoil, as shown already in
Ref.~\cite{DeFazio:2007hw}. In the further work \cite{Gao:2019lta}
it was  demonstrated  that the active light-quark mass corrections can generate  unsuppressed contributions
to the longitudinal $B \to V$ form factors in the heavy quark expansion,
confirming the earlier observation from the power-counting analysis \cite{Leibovich:2003jd}.
The subleading power corrections to these form factors from the two-particle and three-particle HQET DAs
fulfilling the ``classical" operator identities for the light-cone HQET operators
were computed  with the same  twist-six accuracy as for the $B\to P$ form factors.

Importantly, the two-particle twist-five (off-the-light-cone) contributions to the semileptonic $B \to V$ form factors yield sizeable numerical corrections at the level of $(20-30) \%$
of the LP results. Moreover, the long-standing discrepancy between the form-factor ratio
${\cal R}= \left [ (m_B + m_V)  / m_B \right ]  \,  \left [ T_1(q^2) / V(q^2) \right ]$
predicted using the conventional LCSR method with the light vector meson DAs \cite{Ball:2004rg}
and the same ratio obtained from the QCD factorization  \cite{Beneke:2000wa}
has been  clarified in Ref.~\cite{Gao:2019lta},  where a detailed
explanation can be found.

Finally, in Tables \ref{tab:resFFBPBDA}  and \ref{tab:resFFBVBDA} we
collect the most recent predictions for the form factors of $B$-meson transitions
to light mesons at $q^2=0$ obtained from  the SCET sum rules at NLO and from the LCSRs  at LO.
The $q^2$ dependence is given in the papers quoted in these Tables.

%%%%EPJ table
\begin{table}[h]
\begin{tabular}{@{}llll@{}}
%\begin{tabular}{|c|c|c|c|}
%\scriptsize
\toprule
FF & $f^+(0) = f^0(0)$ & $f^T(0)$ & Ref.\\
\midrule
$B \to \pi$ & $ 0.191 \pm 0.073$ & $ 0.222 \pm 0.078$ & \cite{Cui:2022zwm}\\
 & &  &\\[-3mm]
 &$0.21 \pm 0.07$&$0.19 \pm 0.06 $ &\cite{Gubernari:2018wyi}\\
\hline
& &  &\\[-2mm]
$B\to K$    & $0.325 \pm 0.085$  & $0.381 \pm 0.097$ &\cite{Cui:2022zwm}\\
            &  $0.27 \pm 0.08$ & $0.25 \pm 0.07$ & \cite{Gubernari:2018wyi}\\
\hline
& &  &\\[-2mm]
$B_s\to K$ &$0.203 \pm 0.074 $    &$0.260 \pm 0.087$ &\cite{Cui:2022zwm}\\
& &  &\\[-3mm]
\botrule
\end{tabular}
\caption{The LCSR results for $B_{(s)} \to P$ transition form factors at $q^2=0$ from  LCSRs with $B$-meson DAs.
}
\label{tab:resFFBPBDA}%
\end{table}

%%%%EPJ table B\to V
\begin{table}[h]
\begin{tabular}{@{}llll@{}}
%\begin{tabular}{|c|c|c/c/}
%\scriptsize
\toprule
FF (Ref.~\cite{Gao:2019lta})  & $V(0)$            & $A_0(0)$          &  $T_1(0)=T_2(0)$  \\
   &                            & $A_1(0)$          &  $T_{23}(0)$  \\
   &                            & $A_{12}(0)$       &               \\
\midrule
$B\to \rho$ &$0.327_{-0.135}^{+0.204}$&$0.317_{-0.102}^{+0.129}$ &$0.287_{-0.118}^{+0.180}$\\
&&&\\[-2mm]
            &                         &$0.249_{-0.103}^{+0.155}$ &$0.711_{-0.250}^{+0.356}$\\
&&&\\[-2mm]
            &                         &$0.265_{-0.086}^{+0.107}$ &                          \\
&&&\\[-2.5mm]
\hline
&&&\\[-2mm]
$B\to \omega$&$0.357_{-0.148}^{+0.223}$&$0.344_{-0.107}^{+0.142}$&$0.312_{-0.129}^{+0.197}$\\
&&&\\[-2mm]
             &                         &$0.270_{-0.111}^{+0.170}$&$0.767_{-0.266}^{+0.407}$\\
&&&\\[-2mm]
             &        &  $0.284_{-0.087}^{+0.116}$ &  \\
&&&\\[-2mm]
\hline
&&&\\[-2mm]
$B\to K^*$  &$0.419_{-0.157}^{+0.245}$&$0.382_{-0.109}^{+0.154}$  &$0.361_{-0.135}^{+0.211}$\\
&&&\\[-2mm]
            &               &$0.306_{-0.115}^{+0.180}$  &$0.793_{-0.258}^{+0.402}$\\
&&&\\[-2mm]
            &               &$0.273_{-0.077}^{+0.112}$  &                    \\
&&&\\[-2mm]
\hline
\hline
&&&\\[-2mm]
FF (Ref.~\cite{Gubernari:2018wyi})& $V(0)$            & $A_1(0)$          &  $T_1(0)=T_2(0)$  \\
   &                                                   & $A_{2}(0)$        &  $T_{23}(0)$  \\
  \midrule
$B\to \rho$                      &$0.27 \pm 0.14$ &  $0.22 \pm 0.10$ & $0.24 \pm 0.12$\\
            &                         &$0.19 \pm 0.11$        &$0.56 \pm 0.15$\\
\hline
$B\to K^*$  &$0.33 \pm 0.11$  &  $0.26 \pm 0.08$&   $0.29 \pm 0.10$\\
            &               &$0.24 \pm 0.09$& $0.58 \pm 0.13$\\
&&&\\[-4mm]
 \botrule
\end{tabular}
\caption{
The results for $B \to V$ transition form factors at $q^2=0$ from
LCSRs with $B$-meson DAs. }
\label{tab:resFFBVBDA}%
\end{table}

\subsection{Obtaining LCSRs for the $B\to 2\pi$ and $B\to  K\pi$ form factors
}
\label{ssect:B2piKpiBDA}

As already explained  in Section~\ref{ssect:B2piDA}, an accurate
description of semileptonic $B$ decays into broad resonances
such as $\rho(770)$ or $K^*(892)$ demands a calculation of the more general
$B\to 2\pi$ or, respectively, $B\to K\pi$ form factors.
Since the  method of LCSRs with dipion DAs presented in Section~\ref{ssect:B2piDA}
is not yet accurate enough and since the DAs of the $K\pi$ dimeson state are currently not available, an alternative is to use the LCSRs  with $B$-meson DAs. With this method, it is possible to choose an interpolating light-quark current for an arbitrary state
of two pseudoscalar mesons with  a given spin-parity and flavour.
The first such LCSRs  were derived in Ref.~\cite{Cheng:2017smj}
for the $\bar{B}^0\to \pi^+\pi^0$ form factors.
The same correlator was used as the one in Eq.~(\ref{eq:corrB})  introduced originally
\cite{Khodjamirian:2006st} for the
$B\to \rho$ form factor. The idea was to replace  the narrow $\rho$
resonance approximation in the channel of interpolating current by a more general intermediate state of two pions staring from the
threshold $4m_\pi^2$. The $\pi\pi$ state was inserted in the spectral density and the
quark-hadron duality interval $s_0^{2\pi}$ was determined separately, employing
the two-point QCD sum rule with  the same interpolating currents.

For the  $P$-wave $\bar{B}^0\to \pi^+\pi^0$ form factor of the vector weak current defined in Eq.(\ref{eq:BpipiFF}) the resulting sum rule, after applying duality approximation and Borel transform, has the following expression:
\ba
\hspace{-1mm}\int\limits_{4m_{\pi}^2}^{s_0^{2\pi}}\!\!\! ds e^{-s/M^2} \!\!\!\kappa_\perp(s,\!q^2)F^\star_{\pi}(s)F_{\perp}^{(\ell=1)}(s,\!q^2)
\!=\! f_B m_B\!\!\!\!\!\!\!\int\limits_0^{~\sigma_0^{2\pi}(s_0^{2\pi})}\!\!\!\!\!\!\!\!d\sigma e^{-s(\sigma,q^2)/M^2}\
\!\!\frac{\phi_+^B(\sigma m_B)}{1\!-\! \sigma} \!+ \!\dots,
\label{eq:lcsrFperp}
\ea
where $\kappa_\perp(s,q^2)$ is a kinematical factor, $F_\pi(s)$ is the pion
e.m. form factor in the timelike region. On r.h.s. of the above relation
only the leading power contribution of the twist-2 $B$-meson  DA is shown, and the ellipsis denotes power suppressed contributions of higher twist DAs, including the three-particle ones.
Since the above sum rule is obtained only at LO, a default scale $\mu\sim M$ is implied for
the twist-2 DA.
Note that in Eq.~(\ref{eq:lcsrFperp}) the form factor $F_{\perp}^{(\ell=1)}$  is integrated over the dipion invariant mass. Hence, its direct calculation
is only possible in the approximation of a single narrow $\rho$-resonance, in which
case the LCSR for the $B\to \rho$ form factor already considered in Section~\ref{ssect:Blight} is simply restored.
To study the  effects of the $\rho$ width and to assess the role of the nonresonant
background, in Ref.~\cite{Cheng:2017smj} different representations  for the $B\to 2\pi$ form factors with $\ell=1$
 with various resonance content were substituted in the sum rule
 (\ref{eq:lcsrFperp}), and the parameters were fitted to the r.h.s.
 calculated from OPE. For the pion e.m. form factor a data-driven resonance representation was used. The main outcome of the numerical analysis
was  that the cumulative effect of the $\rho$ width and of the nonresonanct
contributions can alter the $B\to \rho$ form factor calculated in the narrow $\rho$-width
approximation by an appreciable   $(15-20) \%$ correction.
The results are also consistent with the ones from Ref.~\cite{Hambrock:2015aor}
(see section~\ref{ssect:B2piDA}) where dipion DAs
were used in the LCSR for the same form factor.

The LCSR method was also applied in
Ref.~\cite{Descotes-Genon:2019bud} and in Ref.~\cite{Descotes-Genon:2023ukb} for the $B\to K\pi$ form factors
with the $K\pi$ state, respectively  in the $P$- and $S$-wave.
These form factors are the most important hadronic inputs
for the analysis of the FCNC $B\to K\pi\ell^+\ell^-$
decays. Note that for the $S$-wave case
in Ref.~\cite{Descotes-Genon:2023ukb}
a nontrivial data-driven representation  of the
$K\pi$ scalar form factors was  used in the sum rule relations, whereas in the $P$-wave case
it was sufficient to use a simple overlap of Breit-Wigner resonances.

\subsection{ The $B$-meson transitions to charmed mesons}
\label{ssect: BDstLCSR}

The $B\to D$ and $B\to D^*$ form factors essential for the observables
in $B\to D^{(*)}\ell\nu_\ell$ decays  are among the best studied
hadronic objects in heavy flavour physics, due to powerful HQET methods allowing one to reduce these form factors to universal Isgur-Wise functions. In addition,
advanced lattice QCD computations  of these form factors are available. However, the  region of momentum transfers beyond the zero-recoil point remains largely
unexplored and relies on the model-dependent extrapolations of HQET form factors. Reliable estimates of power-suppressed $\sim 1/m_c$ effects in this region
are therefore needed.

The method of LCSRs  with $B$-meson DAs is well
suited for the  calculation of the $B\to D^{(*)}$ form factors
in the large recoil region,
as shown first in Ref.~\cite{Faller:2008tr}.
Here again we benefit from the universality of this method.
Replacing in the underlying
correlator (\ref{eq:corrB})  the light quark by the $c$-quark in
both weak and interpolating currents, we repeat all main steps of LCSR
derivation and arrive at the sum rule for a certain $B\to D$ or $B\to D^{*}$ form factor.
A  possibility to assess the $1/m_c$ effects is opened
up by expanding the LCSRs in powers of the inverse $c$-quark mass and comparing the results with the
HQET form factors.

In Ref.~\cite{Gubernari:2018wyi} the LCSRs for the $B\to D$ and
$B\to D^*$ form factors were  further improved and updated
taking into account a more complete  set  of higher-twist $B$ meson DAs
established in Ref.~\cite{Braun:2017liq}.
Furthermore, in Refs.~\cite{Wang:2017jow,Cui:2023bzr}  the SCET sum rules
with $B$-meson DAs  were applied to the $B\to D^{(*)}$ transition form factors
and NLO corrections were computed.  In addition in  Refs.~\cite{Gao:2021sav,Cui:2023bzr}
a variety of the subleading power contributions to these form factors including the higher-twist
corrections have been taken into account in these sum rules. The LCSR results were used
for the $|V_{cb}|$ determination and for predicting the LFU sensitive ratios
${\cal R} (D_{(s)}^{(\ast)} )$.
%
%\begin{table}
%\label{tab:BD}
%	\centering
%	\scriptsize
%	\renewcommand{\arraystretch}{2}	
%	\begin{tabular}{|c|c|c|c|c|c|c|}
%		\hline
%		\hline
%		 $f_{+}^{BD}$ & $f_{0}^{BD}$ & $V^{BD^{*}}$ & $A_{0}^{BD^{*}}$ & $A_{1}^{BD^{*}}$ & %$A_{2}^{BD^{*}}$ & Ref. \\
	%	\hline
%		$0.586(103)$ & $0.586(103)$ & $0.703(160)$ & $0.623(112)$ & $0.704(119)$ & $0.803(186)$&   %\cite{Cui:2023bzr}  \\
%		\hline
 % $0.65 \pm 0.08$   &  --  & $0.69 \pm 0.13$ & -- & $0.60 \pm 0.09$ &$0.51 \pm 0.09$ &  %\cite{Gubernari:2018wyi}  \\
	%	\hline
	%	 $f_{+}^{B_{s}D_{s}}$ & $f_{0}^{B_{s}D_{s}}$ & $V^{B_{s}D^{*}_{s}}$ & $A_{0}^{B_{s}D^{*}_{s}}$ & %$A_{1}^{B_{s}D^{*}_{s}}$ & $A_{2}^{B_{s}D^{*}_{s}}$ & Ref.\\
%		\hline
	%%	$0.572(115)$ & $0.572(115)$ & $0.671(168)$ & $0.582(120)$ & $0.661(117)$ & $0.765(163)$ &  %\cite{Cui:2023bzr}  \\
	%	\hline	
 % \hline
%	\end{tabular}
%	\renewcommand{\arraystretch}{1.0}
%	\caption{ $B_{(s)}\to D_{(s)}^{(*)}$
 %form factors at $q^2=0$ from the LCSRs with $B$ meson DAs. }
%\end{table}
%

%%%%EPJ table B\to charm
\begin{table}[h]
\begin{tabular}{@{}lllll@{}}
%\begin{tabular}{|c|c|c/c/c/}
%\scriptsize
\toprule
FF   & $f^+(0)=f^0(0)$  &    V(0)      & $A_0(0)$    & Ref.      \\
      &                  &              & $A_1(0)$    &           \\
      &                   &                & $A_2(0)$    &           \\
\midrule
$B\to D^{(*)}$ & $0.586\pm 0.103$ & $0.703\pm 0.160$ & $0.623 \pm 0.112$ & \cite{Cui:2023bzr}  \\
          &             &                            & $0.704\pm 0.119$ & \\
          &             &                            &  $0.803 \pm 0.186$ &            \\
          &&&&\\[-1mm]
          &  $0.65 \pm 0.08$   & $0.69 \pm 0.13$ & ~~~~~~-- & \cite{Gubernari:2018wyi} \\
         &                     &                 & $0.60 \pm 0.09$ & \\
         &&                                        &$0.51 \pm 0.09$ &   \\
\hline
$B_s\to D_s^{(*)}$ & $0.572 \pm 0.115$  & $0.671 \pm 0.168$ & $0.582 \pm 0.120$ &\cite{Cui:2023bzr}\\
                   &                      &                  & $0.661 \pm 0.117$  &  \\
                   &                     &                   & $0.765  \pm 0.163$ &  \\
\botrule
\end{tabular}

\caption{ $B_{(s)}\to D_{(s)}^{(*)}$
 form factors at $q^2=0$ from the LCSRs with $B$-meson DAs.}
 \label{tab:resFFBcharmBDA}
 \end{table}

In Table~\ref{tab:resFFBcharmBDA} we collect the $B_{(s)}\to D^{(*)}_{(s)}$ form factors
from the most recent analysis of SCET sum rules from Ref.~\cite{Cui:2023bzr} and compared them
with the results \cite{Gubernari:2018wyi}
of conventional LCSRs at LO and with a different treatment of higher-twist effects .

The method of LCSRs with $B$-meson DAs was also recently used to calculate the form factors of
the semileptonic $B$ decays into charmed
axial $D_1^*$ and scalar $D_0^*$ mesons, in Refs.
\cite{Gubernari:2022hrq} and \cite{Gubernari:2023rfu}, respectively. The derivation of these sum rules, being straightforward on the OPE side (due to a simple replacement of the
spin-parity of the interpolation current), turned out nontrivial
on the hadronic side. In the case of the $B\to D_1^*$ form factors,
a specially designed  procedure with a second interpolation current was needed,
because the two lowest charmed axial mesons have almost the same mass.
Note that, as explained and taken into account in  Ref.~\cite{Gubernari:2023rfu},  for the $B\to D_0^*$ transition, one currently faces a problem to choose between two alternative
 mass patterns of these mesons inferred from the data.

%%%%%%%%%%%%%%%%%%%%%%%%%%%%%%%%%%%%%%%%
\section{Radiative leptonic $B$ decay and related channels}
\label{sect:Bgamlnu}
%%%%%%%%%%%%%%%%%%%%%%%%%%%%%%%%%%%%%%%%
\subsection{ Power suppressed effects in $B \to \gamma \ell \bar \nu_{\ell}$ }

At a large energy of the photon, the hadronic amplitude of the radiative leptonic
$B \to \gamma \ell \bar \nu_{\ell}$ decay
is described by a well studied factorization formula in QCD
\cite{Korchemsky:1999qb,Descotes-Genon:2002crx}, reformulated in SCET \cite{Lunghi:2002ju,Bosch:2003fc}. The LCSR methods play an important
role in this analysis, quantifying the power corrections to the form factors of this decay. The inverse moment $\lambda_B$ of the
$B$-meson DA, discussed in the previous section, directly enters
the LP of the factorization formula in the heavy $b$-quark limit. Therefore,
a measurement of the $B \to \gamma \ell \bar \nu_{\ell}$ decay width anticipated
at Belle-II \cite{Belle:2018jqd}, will provide us with an accurate value of  this
nonperturbative parameter.

The hadronic part of the
$B \to \gamma \ell \bar \nu_{\ell}$
amplitude is described by the nonlocal $B$-to-vacuum matrix element:
\begin{eqnarray}
T_{\mu \nu}(p, q) = -i\int d^4 x \, e^{i p \cdot x}  \,
\langle 0 | {\rm T} \{j_\mu ^{\rm{em}}(x),\,
\bar u(0) \gamma_{\nu} (1-\gamma_5) b  (0) \} |  B(p+q) \rangle
\nonumber\\
= \epsilon_{\mu \nu \rho \sigma} \,
p^{\rho}  v^{\sigma}  F_V(n \cdot p)
+ i\left [ -g_{\mu \nu} v \cdot p + v_{\mu}  p_{\nu} \right ]
F_A(n \cdot p) +\dots\,,
\label{eq:hadtens}
\end{eqnarray}
where $j_\mu^{\rm{em}}$ is the quark
electromagnetic current coupled to a real photon with momentum $p$ and correlated with the weak $b\to u$ current with momentum $q$ transferred to the lepton pair. The relevant part of this hadronic tensor contains the two $B\to \gamma$ form factors $F_V$ and $F_A$. Here  the variable
$q^2$ is customarily replaced with
$n\cdot p=2E_\gamma$, and $E_\gamma$
is the photon energy in the $B$-meson rest frame.
The light-cone vectors are chosen such that
$\bar{n}\cdot p=0$, and $q^2=m_B \, (m_B - n\cdot p)$.

The soft-collinear factorization formula for the  $B \to \gamma$ form factors,
at LP of the expansion in $\Lambda_{\rm QCD}/m_b$, can be cast in the compact  form, equal for both form factors:
\begin{eqnarray}
F_{V, \, A}(n\! \cdot \!p) \equiv F_{\rm LP}(n\!\cdot\! p)= {Q_u \, m_B \over n \!\cdot\! p}  \tilde{f}_B(\mu)  \,
{\cal C}_{\perp}(n\! \cdot\! p, \mu) \!\! \int\limits_0^{ + \infty} \!\!  { d \omega \over \omega} \,
{\cal J}_{\perp}(n \!\cdot\! p,\omega,  \mu) \, \phi_B^{+}(\omega, \mu)  \,, \hspace{0.5 cm}
\label{eq:LP}
\end{eqnarray}
where the $u$-quark charge $Q_u$ indicates that a photon emission from the light quark
dominates the decay amplitude.
At the tree level (LO) the product of the hard and jet functions
in Eq.~(\ref{eq:LP}) is  $[{\cal C}_{\perp}{\cal J}_{\perp}]_{\rm LO}=1$, and the integral reduces to
the desired inverse moment $\lambda_B$.
 The one-loop expressions for the hard function ${\cal C}_{\perp}$ \cite{Bauer:2000yr} and for the jet function ${\cal J}_{\perp}$
\cite{Lunghi:2002ju,Bosch:2003fc}  were obtained with the standard SCET technique.
The most advanced analysis in Ref.~\cite{Beneke:2011nf}  includes RG resummation of the enhanced logarithms of $m_b / \Lambda_{\rm QCD}$ at the NLL accuracy.

However, the power-suppressed ``soft overlap" contributions
to the  $B \to \gamma$ form factors
cannot be estimated within the perturbative factorization framework.
To solve this task, a  QCD-based method,
combining hadronic  dispersion relation with OPE and LCSR in terms of $B$-meson DAs
was suggested in Ref.~\cite{Braun:2012kp}, following the technique
originally developed for the $\gamma^{\ast} \gamma \to \pi^{0}$ form factor in Ref.~\cite{Khodjamirian:1997tk}
(see also Ref.~\cite{Agaev:2010aq} for a further development).

The main idea is  to consider the  $B$-to-vacuum hadronic matrix element
(\ref{eq:hadtens})  at  spacelike $p^2<0$, so that in terms of SCET the electromagnetic current carries a hard-collinear four-momentum.
It is then straightforward to derive factorized expressions of the generalized $B\to \gamma^{\ast} $ form factors $F^{B \to \gamma^{\ast}}_{V,A}( n\!\cdot\! p,p^2)$  in the LP approximation \cite{Wang:2016qii}:
\begin{eqnarray}
F_{V, \, A}^{B \to \gamma^{\ast}} ( n\!\cdot \!p,p^2) &\equiv& F_{\rm LP}^{B \to \gamma^{\ast}} ( n\!\cdot\! p,p^2)
\nonumber\\
&=&
Q_u \, m_B \, \tilde{f}_B(\mu)  \,
{\cal C}_{\perp}(n \!\cdot\! p, \mu) \!
\int_0^{\infty} \!\! d \omega \,  \frac{ {\cal J}_{\perp}(n \!\cdot\! p, p^2 , \omega, \mu) \,  \phi_B^{+}(\omega, \mu)}{(n\!\cdot \!p)\omega - p^2} \,.
\hspace{0.5  cm}
\label{eq:factBst}
\end{eqnarray}
On the other hand, both form  factors obey an unsubtracted hadronic dispersion relation
in the variable $p^2$.  Taking the vector form factor as an example,  this relation reads:
\begin{eqnarray}
F_V^{B \to \gamma^{\ast}}(n\cdot p,\, p^2)
= \frac{f_{\rho} \, F_{B\to \rho}(n\cdot p)}{m_{\rho}^2-p^2}
+ {1 \over \pi}  \, \int\limits_{s_0}^{\infty} \, d s \,
\frac{{\rm Im}_s \, F_V^{B \to \gamma^{\ast}}(n \cdot p, s )}
{s- p^2},
\hspace{1.0  cm}
\label{eq:FVdisp}
\end{eqnarray}
where the ground-state contributions from $\rho$ and $\omega$ are combined into one resonance term
with the narrow-width approximation and with the assumption $m_{\rho} \simeq m_{\omega}$ \cite{Braun:2012kp}.
The numerator in this term contains the decay constant of $\rho$
and the function   $F_{B\to \rho}$ which is, up to some normalization factor, equal to the usual $B\to\rho$
vector form factor $V^{B\to \rho}(q^2)$.

The rest of the derivation consists of standard elements
of the LCSR technique, already presented in the previous section.
The relation (\ref{eq:FVdisp})
at $p^2<0$ is equated to the result of QCD calculation
$F_{\rm LP}^{B \to \gamma^{\ast}}$ given by
Eq.~(\ref{eq:factBst}), and the latter is
transformed into a form of dispersion integral
with imaginary part
${\rm Im}F_{\rm LP}^{B \to \gamma^{\ast}}(n\cdot p,s)$.
After that quark-hadron duality is used to replace the integral on r.h.s. of dispersion relation
(\ref{eq:FVdisp})  by an integral over the
calculated imaginary part ${\rm Im}F_{\rm LP}^{B \to \gamma^{\ast}}$. Applying then the standard LCSR technique enables us to express the product of  $f_\rho$ and the $B\to\rho$ form factor in terms of the subtracted dispersion integral of the QCD spectral density. Returning to the dispersion
relation  (\ref{eq:FVdisp}), and substituting this LCSR in the resonance term
we notice that the limit $p^2\to 0$  can smoothly be taken, resulting in the desired
$B\to \gamma$ form factor for which the following expression is obtained  \cite{Braun:2012kp}:
\begin{equation}
  F_{V}(n\cdot p) =
\frac{1}{\pi}\int\limits_{0}^{s_0} \frac{ds}{m_\rho^2} \mathrm{Im} F_{\rm LP}^{B\to \gamma^* }(n\cdot p,s)e^{-(s-m^2_\rho)/M^2} +
\frac{1}{\pi}\int\limits_{s_0}^\infty \frac{ds}{s} \mathrm{Im} F_{\rm LP}^{B\to \gamma^* }(n\cdot p,s)\,,
\label{eq:general}
\end{equation}
and analogous expression for the axial form factor $F_A$.
Further analysis of these relations allows one to represent it  as a sum
of the LO contribution and the needed power-suppressed soft overlap
correction. In Ref. \cite{Beneke:2018wjp} this approach was further developed, including higher-twist contributions to $B$-meson DAs.

An alternative LCSR technique to address the power-suppressed effects in the $B\to \gamma \ell \bar \nu_{\ell}$ amplitude
uses a correlation function with the  photon DAs and $B$-meson interpolation current:
\begin{eqnarray}
\widetilde{T}_{\nu}(p, q) = i\int d^4 x \, e^{i q \cdot x}  \,
\langle \gamma(p) | {\rm T} \{
\bar u(x) \gamma_{\nu} (1-\gamma_5) b  (x), \,  m_b \, \bar{b}(0)i\gamma_5 u(0) \} | 0 \rangle \,.
\label{eq:gamm}
\end{eqnarray}
The photon DAs emerging after factorizing this correlator describe
long-distance photon emission, the so called hadronic component of the photon.
This approach should be put in one category with the LCSRs using light-meson DAs,
so that here a central role is played by a twist expansion
of photon DAs worked out in Ref.~\cite{Ball:2002ps}.  A key nonperturbative parameter in the
leading-twist photon DA  is the magnetic susceptibility of the quark condensate
\cite{Ioffe:1982qb} which describes the response of the QCD vacuum to
an external electromagnetic field. A specific feature of the LCSRs with photon DAs  is that,
apart from the long-distance photon emission encoded in these DAs,
the photon emission at short distances described by triangle heavy-light diagrams also contributes to the correlator.
Furthermore, these sum rules
are universal with respect to heavy flavour, that is,  it is possible to switch to the charmed quark
correlator with the same OPE and access also the radiative leptonic $D\to \gamma \ell\nu_\ell$ decay.
At leading order the LCSRs with photon DAs were obtained
in the early papers \cite{Khodjamirian:1995uc,Ali:1995uy,Eilam:1995pp}, with an NLO improvement in
more recent works \cite{Ball:2003fq,Wang:2018wfj}.  Technically involved NLO gluon radiative corrections
to the point-like photon contribution were only recently computed in Ref.~\cite{Janowski:2021yvz}.

 The four-body leptonic $B\to \mu \bar \mu \ell \bar \nu_{\ell}$ decay with three charged leptons in the final state
 (free of the  helicity suppression) belongs to the rare
 $B$-meson decay channels accessible at hadron collider, as opposed to  the two-body leptonic decay.
 The hadronic part of the
 $B\to \mu \bar \mu \ell \bar \nu_{\ell}$
 decay amplitude, albeit formally described by the
 same Eq.(\ref{eq:hadtens}), where the real photon is replaced with a virtual one, is in fact
 far more complicated than for $B\to \gamma \ell \bar \nu_{\ell}$ because: (i) there are now  three independent form factors,
 (ii) the timelike photon with $p^2>0$ emitted from the light quark
 generates intermediate hadronic states dominated by resonances with $\rho,\omega$ meson quantum numbers.
 A factorization pattern for this hadronic amplitude is valid
if the photon is spacelike,
with a virtuality of at least ${\cal O}(m_b\Lambda_{\rm QCD})$
 (a usual hard-collinear scale). In Ref. \cite{Wang:2021yrr},
applying the two-step matching  ${\rm QCD} \to {\rm SCET_{I}} \to {\rm SCET_{II}}$
for the hadronic tensor, the factorized expressions for the off-shell $B \to \gamma^{\ast} $ form factors
at the LP accuracy were obtained, including also the NLL resummation of the parametrically enhanced logarithms.
Subsequently, the power-suppressed contributions from four distinct sources have been computed with the same factorization method at tree level. The results respect the constraints on
the $B \to \gamma^{\ast} W^{\ast}$ form factors due to QED gauge invariance of the electromagnetic interaction
(see also Ref.~\cite{Beneke:2021rjf}). The use of these form factors for various decay observables at
large timelike $p^2$, above ground-state resonances, provides to  a usual QCD factorization approximation.
A complementary study of this process from the hadronic side was also done
recently in Ref.~\cite{Kurten:2022zuy}, employing the vector-meson dominance ansatz and the $z$-series parametrization for the off-shell $B \to \gamma^{\ast}$ form factors.
A more systematic approach combining the factorization formulas with dispersion relations
and the  elements of LCSR technique is a perspective future task.

\subsection{ Radiative decays of  the heavy vector mesons from LCSRs }

 To access these decays, we consider  the  vacuum-to-photon correlator
 (\ref{eq:gamm}) and retain only
 the vector $b\to u$ current, considering it as  an  interpolation
 current for the $B^*$ meson. Matching this correlator to a double dispersion relations in both $B^*$ and $B$ channels yields
 LCSRs for the $B^{\ast} B \gamma$ and -- after $b\to c$ replacement --
for the  $D^{\ast} D\gamma$ ``magnetic" couplings determining the radiative decays
$B^{\ast}\to B \gamma$  and $D^{\ast} \to D\gamma$. An early application of these sum rules can be found in Ref.~\cite{Aliev:1995zlh}. A more elaborated LCSR computation of these couplings at LO and in the twist-four approximation was accomplished  in Ref.~\cite{Rohrwild:2007yt}. The result was used
to extract the value of the magnetic susceptibility
from the measured branching fraction ${\cal BR}(D^{\ast 0} \to D^0 \gamma)$,
confirming an earlier determination with the method of QCD sum rules \cite{Ball:2002ps}.
Recently, the NLO QCD corrections to the hadronic photon contribution at twist-two
in the LCSRs were computed in Ref.~\cite{Li:2020rcg}.
In  Ref.~\cite{Pullin:2021ebn}, the NLO QCD corrections to the short-distance photon
emission diagrams
in these sum rules  have been calculated. In Table~\ref{tab:BstBgam}, we present the two most recent LCSRs predictions. Note that, in contrast to Ref.~\cite{Pullin:2021ebn},
the  complete set of photon DAs at the twist-four accuracy \cite{Ball:2002ps} was employed in Ref.~\cite{Li:2020rcg}, where one can
find a detailed discussion and comparison with other theory predictions.
%
%\begin{table}[t]
%\scriptsize
%\begin{center}
%\renewcommand{\arraystretch}{1.5}
%\renewcommand{\multirowsetup}{\centering}
%\resizebox{\columnwidth}{!}{
%\begin{tabular}{|c||c|c|c||c|c|c|}
%\hline
%\hline
 % & $g_{D^{\ast +}  D^+  \gamma}$ & $g_{D^{\ast 0} D^{0} \gamma}$  &  $g_{D_s^{\ast +}D_s^{+} \gamma}$
%  & $g_{B^{\ast +} B^{+} \gamma}$ & $g_{B^{\ast 0} B^{0} \gamma}$  &  $g_{B_s^{\ast 0} B_s^{0} \gamma}$  %\\
%& $({\rm GeV}^{-1})$  & $({\rm GeV}^{-1})$  &  $({\rm GeV}^{-1})$
%&  $({\rm GeV}^{-1})$ & $({\rm GeV}^{-1})$  &  $({\rm GeV}^{-1})$  \\
%\hline
%NLL LCSR \cite{Li:2020rcg} & $-0.15^{+0.11}_{-0.10}$  & $1.48^{+0.29}_{-0.27}$ & %$-0.079^{+0.086}_{-0.078}$
%& $1.44^{+0.22}_{-0.20}$ & $-0.91^{+0.12}_{-0.13}$ & $-0.74^{+0.09}_{-0.10}$ \\
%\hline
%NLO LCSR \cite{Pullin:2021ebn} & $-0.40^{+0.12}_{-0.13}$ & $2.11^{+0.35}_{-0.34}$ & %$-0.60^{+0.19}_{-0.18}$ & $1.44^{+0.27}_{-0.26}$
%& $-0.86 \pm 0.15$ & $-0.95^{+0.15}_{-0.16}$  \\
%\hline
%\end{tabular}
%}
%\end{center}
%\vspace{0.2cm}
%\caption{The most recent LCSR predictions of the couplings  $H^{\ast} H \gamma$ ($H=D_{(s)}, \, %B_{(s)}$)}
%\label{tab:HstHgam}
%\end{table}

%%%%EPJ table
\begin{table}[h]
\begin{tabular}{@{}lll@{}}
%\begin{tabular}{|c|c|c|}
%\scriptsize
\toprule
coupling & Ref.~\cite{Li:2020rcg} (NLL)  & Ref.~\cite{Pullin:2021ebn} (NLO)\\
\midrule
$g_{B^{\ast +} B^{+} \gamma}$ & $~~1.44^{+0.22}_{-0.20}$ & $~~1.44^{+0.27}_{-0.26}$\\
 & &  \\[-2mm]
 \hline
 &&\\[-2.5mm]
 $g_{B^{\ast 0} B^{0} \gamma}$  &$-0.91^{+0.12}_{-0.13}$ &$-0.86 \pm 0.15$\\
&   &\\[-2.5mm]
\hline
&   &\\[-2.5mm]
$g_{B_s^{\ast 0} B_s^{0} \gamma}$   & $-0.74^{+0.09}_{-0.10}$& $-0.95^{+0.15}_{-0.16}$\\
\botrule
\end{tabular}
\caption{The LCSR predictions of the  $B^{\ast} B \gamma$ coupling.}
\label{tab:BstBgam}%
\end{table}

\subsection{ $B_{s,d} \to \gamma \gamma$ and
$B_{s,d} \to \mu \bar \mu\gamma$ decays    }

Among various rare decays of $B_{s(d)}$ mesons  mediated by the FCNC $b\to s(d)$
transitions, the  double radiative $B_{s(d)} \to \gamma \gamma$ decays, despite
a seemingly simple non-hadronic final state, are quite complicated
processes from the QCD point of view,  with a rich hierarchy
of effective operators and several contributing quark topologies.
The LP contributions to the two helicity form factors of $B_{s,d} \to \gamma \gamma$
have been determined with the QCD factorization approach at ${\cal O}(\alpha_s)$ \cite{Descotes-Genon:2002lal} but without including the two-loop $b \to q \gamma$ matrix elements of QCD penguin operators \cite{Buras:2002tp,Asatrian:2004et}.
Factorization properties of the power-suppressed weak annihilation contributions
stemming from the current-current operators were explored at two loops in Ref.~\cite{Bosch:2002bv}, where  the one-loop short-distance functions were also obtained.

A complete NLL computation of the $B_{s, d} \to \gamma \gamma$ decay amplitudes at the LP accuracy
has been  performed in \cite{Shen:2020hfq},
by employing the two-loop RG evolution equation of $\phi_B^{+}(\omega, \mu)$ \cite{Braun:2019wyx}.
In the same paper, subleading-power contributions  from five distinct dynamical sources
have been established at tree level in a factorized form, expressed via the two- and three-particle
$B$-meson DAs of higher twist.
However, as  explained in detail in Ref.~\cite{Shen:2020hfq}, the power-suppressed ``resolved" photon contribution  to the $B_{s,d} \to \gamma \gamma$ amplitude
cannot be computed with the SCET factorization formalism. Here is where the
combination of OPE, LCSRs and dispersion relations
discussed in the context of the $B \to \gamma \ell \bar \nu_{\ell}$ decay, enters the stage. The following correlator of the effective operator $O_7$ and
the electromagnetic current with a  virtual momentum $q^2<0$ was considered:
\begin{eqnarray}
\tilde{T}^7_{\alpha \beta}(p,q) =
2 \overline{m}_b \!\!
\int \!\!d^4 x e^{i q \cdot x} \langle 0 | {\rm T} \!\left \{ j^{\rm em}_{\beta}(x),
\bar q_L(0) \sigma_{\mu \alpha} p^{\mu} b_R(0) \right \}\!|\bar{B}_{q} \rangle
+  \left [ p \leftrightarrow q, \alpha \leftrightarrow \beta \right ].
\hspace{0.2 cm}
\label{eq:deftens}
\end{eqnarray}
From  the results obtained in Ref.~\cite{Shen:2020hfq}, we only
mention that the power-suppressed soft contribution reveals a destructive interference with the  LP effect.
Another type of the subleading power contribution to $B_{s, d} \to \gamma \gamma$ from the soft gluon radiation off the  quark loop
has been recently computed \cite{Qin:2022rlk} in terms of a factorization approach,
 with a generalized soft function defined by the HQET matrix element of the non-local operator
with quark-gluon fields localized on different light-cone directions.

Even more  important are the exclusive FCNC
$B_{d, s} \to \mu \bar \mu\gamma$ decays.
In particular, the angular distributions of these radiative leptonic decays provide
interesting observables  linked to the effective couplings
in the weak effective $b \to q  \ell \bar \ell$ Lagrangian. Albeit with an additional suppression by the QED coupling $\alpha_{\rm em}$,
the  $B_{d, s} \to \mu \bar \mu\gamma$ widths do not suffer from the helicity suppression
inherent to  the purely leptonic $B_{d, s} \to \mu \bar \mu$ decays.

A systematic investigation of the  $B_{d, s} \to \mu \bar \mu\gamma$  form factors
with an energetic photon has been carried out at LP in the heavy quark expansion
with the SCET factorization technique \cite{Beneke:2020fot},
thus going beyond previous works based upon  model-dependent approximations.
In the same paper, the power-suppressed corrections were also computed
stemming from: (i) the  photon radiation off the heavy quark, (ii)
the subleading terms in the expansion of the hard-collinear quark propagator,
and (iii) the weak-annihilation diagrams with insertions of the four-quark operators.
However, similar to the situation with $B\to \gamma \ell \bar \nu_\ell$ discussed above, there are two power-suppressed soft form factors,
which were pragmatically treated with the  resonance\! $\oplus$ \!factorization ansatz.
Evaluating these contributions to the off-shell $B_q(p_B) \to \gamma^{\ast}(q) \, \gamma(k)$
form factors in the time-like $q^2$ region with the help of an OPE-controlled dispersion relation  remains an interesting  task for the future.

\

%%%%%%%%%%%%%%%%%%%%%%%%%%%%%%%%%%%%%%%%
\section{A brief guide to other applications }
%%%%%%%%%%%%%%%%%%%%%%%%%%%%%%%%%%%%%%%%
\label{sect:other}

\subsection{ Form factors of $\Lambda_b\to$ baryon transitions from LCSRs }
\label{ssect:Lambdab}
Apart from many applications to $B$-meson decays, the method of LCSRs  was also  used  to obtain
the form factors of semileptonic heavy-baryon decays at large hadronic recoil. Both versions of LCSRs were applied, based, either (I) on the light-baryon (nucleon or strange hyperon) DAs
 or (II) on the heavy baryon DAs, the latter defined in HQET.

With the  first version of the method, the complete set of the  $\Lambda_b \to p$ form factors
was computed in Ref.~\cite{Khodjamirian:2011jp}. The vacuum-to-nucleon correlator (in the isospin symmetry limit)
\begin{equation}
\Pi_a(P,q)=i\int d^4z\ e^{iq\cdot
z}\langle 0 |T\left\{\eta_{\Lambda_b}(0),j_a(z)\right\}|N(P)\rangle
\label{eq:corrN}
\end{equation}
was used, where $j_a$ is the weak $b\to u$ transition current and $\eta_{\Lambda_b}$ is the interpolation current of the $\Lambda_b$
baryon. For this correlator, the OPE in terms of nucleon DAs was obtained, achieving
the twist-6 level.  These  DAs  were worked out in, e.g.,  Refs.~\cite{Braun:2000kw,Braun:2006hz}, aimed
at the studies of LCSRs for the nucleon electromagnetic form factors.

In Ref.~\cite{Khodjamirian:2011jp},  the correlator  (\ref{eq:corrN}) was matched to
the hadronic dispersion relation in the variable $(P-q)^2$ and, accordingly, the quark-hadron duality
in the $\Lambda_b$ channel was applied.
One of the  novelties suggested in that work, was the procedure  to eliminate
the unwanted ``contamination" from the contributions of negative-parity heavy baryons in the
resulting sum rules.
The advantage of the version I for baryonic LCSRs  is the possibility to easily switch to the charm sector by a $b\to c$ replacement in the correlator. In this way,  phenomenologically important byproducts --
the strong couplings
$\Lambda_c N D^{(\ast)}$ and $\Sigma_c N D^{(\ast)}$ -- have also been computed
in Ref.~\cite{Khodjamirian:2011jp}, applying the technique  with a double dispersion relation,
similar to the one used for  the strong couplings of bottom mesons
and discussed in section~\ref{ssect:BpiLCSRlight}.
Among other applications of this method,
 LCSRs for  the semileptonic $\Lambda_b \to \Lambda$ form factors were obtained in Ref.~\cite{Wang:2008sm}, employing the  $\Lambda$-baryon DAs \cite{Huang:2006ny} (see Ref.~\cite{Liu:2008yg} for further discussions).

An alternative  version II  for the heavy-to-light baryonic form factors
was first discussed in Ref.~\cite{Wang:2009hra} at LO,
employing  the $\Lambda_b$-baryon DAs in HQET worked out in many details in Ref.~\cite{Ball:2008fw}.
The NLO QCD corrections to the ten independent $\Lambda_b \to \Lambda$ helicity form factors were computed  in the LCSR framework in Refs.~\cite{Feldmann:2011xf,Wang:2015ndk}.
In particular, in Ref.~\cite{Wang:2015ndk} the factorization-scale independence of the $\Lambda_b$-to-vacuum correlator has been verified explicitly at the one loop level. Extending the LCSR technique with the HQET heavy-baryon DAs to the semileptonic  $\Lambda_b \to \Lambda_c$ form factors will also appear soon \cite{Lambdab-to-Lambdac}.

In the future, the accuracy of heavy baryon form factors from LCSRs can be further
increased. In the version  I of the method, it is desirable to improve  our knowledge
of the nucleon and light hyperon DAs (e.g., from lattice QCD).
In the version II based on HQET and SCET technique, one should investigate additional perturbative contributions. For example, it is anticipated \cite{Wang:2011uv,Mannel:2011xg} that
LP contributions to the  heavy-baryon decay form factors arise from the spectator scattering mechanism with  two hard-collinear gluon exchanges. To prove that within the LCSR framework, a computation of the  two-loop diagrams  in the underlying correlator is necessary.

\subsection{Nonleptonic two-body decays of $B$-meson  }

Even the simplest two-body weak nonleptonic decays of $B$ meson, such as $B\to \pi\pi$, are characterized by a rich pattern of
contributions to the decay amplitude with different quark topologies
(emission, exchange, penguin, annihilation etc.). An additional
complication is caused by the hadronic final-state interactions in these decays.
In the current analyses of nonleptonic decays,
QCD is systematically used only in the formation of the
effective Hamiltonian, taking into account virtual gluons at the
energy-momentum scales between $m_W$ and $m_b$.
A complete QCD-based calculation of hadronic amplitudes
relevant for   nonleptonic  $B$ decays
is  considerably more challenging  than for
semileptonic or radiative $B$ decays,  which  are usually fully factorized
into  hadronic transition form factors.
A systematic approach to nonleptonic $B$ decays, known as the QCD factorization (QCDF)
\cite{Beneke:1999br,Beneke:2001ev}, exists in the limit  $m_b\to \infty$. Within QCDF,  two-body nonleptonic  $B$ decays are described  with
a reduced amount of universal hadronic quantities,
such as decay constants, $B$-meson transition form factors, as well as  the
light-meson and $B$-meson DAs. However, despite many successful applications to describe
nonleptonic decay widths, the  CP asymmetries predicted from QCDF
generally differ from the corresponding experimental values, signalling that the
power suppressed effects of $O(1/m_b)$ are important in the phenomenological
applications. A reliable
estimate of these  effects demands a  QCD-based method employing nonperturbative elements and
retaining a finite $b$ quark mass.

LCSRs for the $B\to \pi\pi$  decays were introduced in Ref~\cite{Khodjamirian:2000mi}.
The method is based on a vacuum-to-pion matrix element correlating
the  $B$-meson- and pion-interpolating currents with an operator of the effective
weak Hamiltonian. To avoid ``parasitic" contributions of
light intermediate hadronic states in the $B$-meson channel, an auxiliary momentum
is attributed to the effective vertex. The sum rule is obtained in  three
steps, applying: (i) dispersion relation and duality in the second  pion channel,
(ii) local duality approximation, that is,
a transition from a spacelike value of the final-state invariant mass squared to
its physical timelike value $m_B^2$, and
(iii) dispersion relation  and duality in the $B$ meson channel.
Using  a finite $b$-quark mass allows one to quantify the $O(1/m_b)$ corrections,
reproducing at the same time the QCDF results at $m_b\to \infty$.
In particular, the purely factorizable part is naturally reproduced where
the $B\to \pi$ form  factor is represented by the LCSR with the pion DAs.
In addition, in Ref.~\cite{Khodjamirian:2000mi} a soft-gluon nonfactorizable contribution to the emission
topology in $B\to \pi\pi$ was estimated which is a typical power suppressed effect not accessible
in QCDF.

Subsequent uses of LCSRs for $B\to \pi\pi$ included
the gluonic chromomagnetic operator contribution
\cite{Khodjamirian:2002pk},
as well as the charm penguin \cite{Khodjamirian:2003eq}
and the annihilation topology
\cite{Khodjamirian:2005wn} contributions.
Importantly, the latter is not divergent at a finite $m_b$, as it appears to take place
in QCDF (see also a recent discussion in Ref.~\cite{Lu:2022kos}).

Turning to other applications, the $B$-meson two-body decays into kaon and charmonia  were also
calculated using LCSRs in Refs.~\cite{Melic:2003bw,Melic:2004ud}.
Quite recently, in Ref.~\cite{Piscopo:2023opf} the soft nonfactorizable correction to the
$B\to D\pi$ decay was  obtained using a version of LCSRs with $B$ meson DAs.

Assessing the perspectives of the LCSRs for nonleptonic decays,
we have to take into account that this method has an additional ``systematic" uncertainty caused by applying the local duality approximation
(the step (ii) in the derivation described above in this subsection). Hence, the resulting sum rules  are, in general , less accurate than  LCSRs for the $B$ transition form factors presented in the previous sections.
Still, the method has certain perspectives, mainly because the amplitudes
calculated from LCSRs involve
power suppressed  contributions  not accessible or inherently divergent in QCDF.
A complete analysis of $B$ decays into
two pseudoscalar mesons, such as $B_{(s)}\to K\pi, \bar{K}K$
is one of the perspective applications, especially
in view of certain tensions between data and  QCDF results.

\subsection{Nonlocal effects in $B\to K^{(*)}\ell^+\ell^-$ decays}
\label{ssect:BKelell}
The $B\to K^{(*)}\ell^+\ell^-$ decays generated by
 the $b\to s \ell^+\ell^-$ transitions
with flavour-changing neutral currents (FCNC)
provide sensitive tests of  the flavour sector in SM
(for a recent review, see e.g., Ref.~\cite{Capdevila:2023yhq} in this volume).
LCSRs were used to provide hadronic input in these decays.
For the simplest decay $B\to K \ell^+\ell^-$,  the full decay amplitude
in SM is described by the following formula:
\begin{eqnarray}
&&A (B(p\!+\!q) \to K(p) \ell^{+} \ell^{-}) = { G_F \over  \sqrt{2} }
{\alpha_{em} \over \pi} V_{tb} V_{ts}^{\ast} \Bigg[
\bar{\ell}\gamma_{\mu} \ell\, p^\mu\bigg( C_9 { f^{+}_{BK}(q^2)}
\nonumber \\
&&+ {2 (m_b+m_s) \over m_B+m_K} C_7 { f^{T}_{BK}(q^2) }
+\!\!\!\!\!\!\sum\limits_{i=1,2,...,6,8}\!\!\!C_i
~{\cal H}^{(BK)}_{i}(q^2)
 \bigg)
+ \bar{\ell} \gamma_{\mu} \gamma_5 \ell \, p^\mu C_{10}
{ f^{+}_{BK}(q^2)} \bigg],
\label{eq:ABKelel}
\end{eqnarray}
where genuine FCNC contributions sensitive  to new physics are given by the
terms proportional to the effective coefficients $C_{7,9,10}$. The hadronic parts of these contributions
are  reduced to the $B\to K$ form factors $f_{BK}^{+,T}$ obtained from lattice QCD  or from LCSRs. The remaining terms in Eq.~(\ref{eq:ABKelel})
are expressed  in terms of  hadronic matrix elements
\be
\langle K(p)|i\!\int d^4x\, e^{iqx}\,
 T\{ j_\mu^{em}(x), O_i(0)\}|B(p+q)\rangle = p_\mu { {\cal H}_{i}^{(BK)}(q^2)}\,,
\label{eq:Hnonloc}
\ee
where $O_i$ are the contributing effective operators
\footnote{see e.g.,
Ref.~\cite{Khodjamirian:2012rm} for definitions of  these operators  and their Wilson coefficients.}
and $j_\mu^{em}$
is the quark e.m. current.
The $B\to K^*\ell\ell$ decay has a more rich kinematical structure with three invariant amplitudes
analogous to  Eq.~(\ref{eq:ABKelel}), each of them
containing hadronic contributions similar to the one in Eq.~(\ref{eq:Hnonloc}).

The nonlocal hadronic matrix elements (\ref{eq:Hnonloc})
are known as ``charm loops", since they are
dominated by a weak  $b\to c\bar{c} s$ transition followed by a lepton pair emission via virtual photon: $\bar{c}c\to \gamma^*\to \ell^+\ell^-$. Depending on the momentum transfer $q$, the
intermediate $c\bar{c}$ pair either forms a virtual charm loop (at $q^2\ll 4m_c^2$) or transforms to
an on-shell  charmonium resonances (at $q^2\geq m_{J/\psi}^2$).
In the region far below these resonances, there is still a possibility to use QCD factorization,
as it was done in Ref.~\cite{Beneke:2001at}. However, with this method two problems arise:
the $\bar{c}c$ threshold explicitly enters observables instead of hadronic thresholds  and
the soft gluon exchanges between the charm loop and the rest of the hadronic transition are not accessible.

In Ref.~\cite{Khodjamirian:2010vf}  a new method to obtain the charm loop contributions
in $B\to K^{(*)}\ell^+\ell^-$
was suggested. The idea was to calculate the amplitudes (\ref{eq:Hnonloc}) at spacelike $q^2$,
using light-cone OPE for the charm loop and including the soft-gluon contributions.
For the latter, LCSRs with three-particle $B$-meson DAs were used.
The result of the OPE  was then fitted at $q^2<0$ to a dispersion relation in the $q^2$-variable, containing the  $J/\psi$ and  $\psi(2S)$ poles and a certain ansatz for the integral over the spectral density of excited $\bar{c}c$ hadronic states.  Finally, this relation provides an estimate of nonlocal amplitudes at timelike $q^2$ up to charmonium threshold.

In Ref.~\cite{Khodjamirian:2012rm} a complete calculation of the amplitudes (\ref{eq:Hnonloc})
for $B\to K\ell^+\ell^-$ was done, followed by an update and inclusion of  $B_s$ modes
in Ref.~\cite{Khodjamirian:2017fxg}. In these analyses,  certain contributions were estimated using QCD factorization, since their sum rule calculation is technically not feasible.
More recently, in Refs.~\cite{Gubernari:2020eft,Gubernari:2022hxn}  along with a further update of nonlocal effects in $B\to K \ell^+\ell^-$ and $B_s\to \phi \ell^+\ell^-$decays,
the LCSR for the soft-gluon contribution was recalculated with an updated set of higher-twist  $B$-meson DAs, yielding
a significantly smaller effect.

Note that nonlocal effects in exclusive
FCNC decays are not yet accessible in QCD on the lattice.
Hence OPE- and  LCSR-based calculations combined with hadronic dispersion relations
will remain the only available continuum QCD tool.
This combined approach, however,  still has the following problems deserving dedicated studies:
(i) the calculation in the spacelike $q^2$ region is strictly speaking not a regular OPE, because the
charm loop contribution
even at $q^2\ll 4m_c^2$ is complex valued,  hence contains intermediate  $\bar{c}cs \bar{q}$
hadronic states (where $\bar{q}$ is a spectator quark in the  $B$-meson);
(ii) the usual three-particle DAs of $B$ meson are kinematically not well suited for the
correlators used for estimating soft-gluon contributions (see recent discussion in \cite{Qin:2022rlk}). In this situation, an alternative method
of calculation is desirable, e.g.  the use of LCSRs  with the kaon, $K^*$ and $\phi$  DAs.

\subsection{Decays of $B$ meson into dark matter particles}

The method of LCSRs can be easily extended
to $B$-meson  transition form factors and other hadronic matrix elements emerging
due to new interactions beyond SM.
If a new particle is coupled to quarks and gluons via an  effective pointlike
interaction
(e.g., due to heavy mediators), the corresponding local operator can be inserted in the vacuum-to-hadron correlator
instead of an SM current, resulting in a sum rule for
 the $B$ decay amplitude into a hadron and a new particle.
A recent example is the calculation of the $B$-meson decay
rate into a proton and dark antibaryon in Ref.~\cite{Khodjamirian:2022vta}.
This type of decays is
predicted in the $B$-mesogenesis scenario introduced in Ref.~\cite{Elor:2018twp},
where one can find all necessary details concerning the new physics aspects and the
quark-level interactions with dark particles.

In  Ref.~\cite{Khodjamirian:2022vta} the effective coupling between three quarks and dark antiybaryon
was correlated with the B-meson interpolating current in the vacuum-to-nucleon correlator.
The OPE with nucleon DAs was then used to the leading twist-3 accuracy,
and LCSR was obtained after matching the OPE with the dispersion relation in the $B$-meson channel.
A nontrivial difference is observed between various versions of B-mesogenesis model
in which the effective interaction  differs only by interchange of $b$ and $d$ quarks.
The $B\to $ nucleon  effective form factors calculated from LCSRs
were used to predict the branching ratios of the $B$ decays into a proton and
dark-matter antibaryon. An important upgrade of this calculation  will include
the nucleon DAs up to twist-6  \cite{ABMW}.
Recently, the same LCSR method was used
in Ref.~\cite{Elor:2022jxy} also for the $B$ decay modes into other baryons
and dark antibaryons.
The predicted decay rates  are within the reach of the Belle-II experiment, where a final-state proton (or other baryon) and a missing energy-momentum in the $B$ decay
serve as a signature.

%%%%%%%%%%%%%%%%%%%%%%%%%%%%%%%%%%%%%%%%
\section{Conclusion }
%%%%%%%%%%%%%%%%%%%%%%%%%%%%%%%%%%%%%%%%

The method of light-cone sum rules, after many years of
successful development and plenty of topical applications,
has become a standard QCD-based tool
to calculate hadronic matrix elements for $b$-quark decays
in a form of approximate analytic expressions with an improvable accuracy.
In this review, we presented both
main versions of the LCSR technique, based on the light-meson
and $B$-meson DAs.
In particular, the $B$-meson transition form factors
and other exclusive $B$-decay amplitudes at large recoil of the final state
are calculable with LCSRs, successfully complementing
the lattice QCD calculations.
 A broad variety of other  prospective applications,
from the heavy baryon form factors to the $B$-decays into dark matter
were also overviewed. We collected references to
all essential papers on the LCSR applications to heavy hadron physics. Having in mind the vast amount
of literature in this field, it is possible that some of the relevant publications are still overlooked.

We hope that this review will become useful, especially for  young researchers
who enter the field of QCD methods in flavour physics, providing guidance to
their future works on new interesting applications of QCD light-cone sum rules.

\label{sect:outl}

\section*{Acknowledgements}

The work of A.K. is supported by DFG (German Research Foundation) under grant 396021762-TRR 257 “Particle Physics Phenomenology after the Higgs Discovery”. B.M. acknowledges support from the Alexander von Humboldt Foundation
in the framework of the Research Group Linkage Programme, funded by the German Federal Ministry of Education, as well as the support of the Croatian Science Foundation (HRZZ) under project “Heavy hadron decays and lifetimes” (IP-2019-04-7094).
Y.M.W. acknowledges support from the  National Natural Science Foundation of China  with
Grant No. 11735010 and 12075125, and the Natural Science Foundation of Tianjin
with Grant No. 19JCJQJC61100.

\bigskip

{\bf Data Availability Statement:} No Data associated in the manuscript.

%previous bibtex
%\bibliographystyle{JHEP.bst}
%\bibliography{References_LCSR_review.bib}

%EPJ bibliography
\bibliography{References_LCSR_review}

\end{document}